\documentclass[a4paper,11pt]{article}%
\usepackage{amssymb}
\usepackage{graphicx}
\usepackage{amsbsy}
\usepackage{amsmath}
\usepackage{cite}
\usepackage{slashed}
\usepackage{amsfonts}
\usepackage{hyperref}
\usepackage{comment}
\usepackage{braket}
\usepackage{tensor}
\usepackage{array}

\usepackage[utf8]{inputenc}

\usepackage{tikz}
\usetikzlibrary{shapes,arrows,positioning,automata,backgrounds,calc,er,patterns}
\usepackage[compat=1.1.0]{tikz-feynman} 

\usepackage{listings}
\usepackage{color}

\definecolor{mygreen}{rgb}{0,0.6,0}
\definecolor{mygray}{rgb}{0.5,0.5,0.5}
\definecolor{mymauve}{rgb}{0.58,0,0.82}
\definecolor{darkWhite}{rgb}{0.94,0.94,0.94}

\newcommand{\dd} {\mathrm{d}}

\newcommand{\tr} {\mathrm{tr}}
\newcommand{\Tr} {\mathrm{Tr}}
\newcommand{\boundellipse}[3]
{(#1) ellipse (#2 and #3)
}
\newcommand{\dc} {\mathcal{D}}
\newcommand{\Lagr}{\mathcal{L}}

\newcommand{\sD}{\slashed{D}}
\newcommand{\sdc}{\slashed{\mathcal{D}}}
\DeclareMathOperator{\pa}{\partial}
\DeclareMathOperator{\n}{\nabla}
\DeclareMathOperator{\g}{\sqrt{-\textit{g}}}
\newcommand{\lm}{\log\left(\frac{m^2}{\mu^2}\right)}
\newcommand{\dq}{\frac{\dd^d q}{(2\pi)^d}}

\newcommand\T{\rule{0pt}{3.2ex}}
\newcommand\B{\rule[-2.1ex]{0pt}{0pt}}

\usepackage{dsfont}

\def\logm#1{\log\frac{#1}{\mu^2}}

\usepackage{multicol}
\usepackage{color}
\definecolor{verdes}{cmyk}{0.92,0,0.59,0.4}

\definecolor{Grn}{rgb}{0.1,0.5,0.2}
\definecolor{Blu}{rgb}{0.,0.,1.}
\definecolor{Red}{rgb}{0.7,0.1,0.1}
\definecolor{SE}{rgb}{0.5,0,0.4}
\definecolor{Tur}{rgb}{0,0.75,0.65}

\usepackage{hyperref} 
\hypersetup{
     colorlinks = true,
     linkcolor = blue,
     citecolor = magenta,
     anchorcolor = blue,
     filecolor = blue,
     urlcolor = magenta,
     bookmarks=true,
     linktocpage =true,
     }

\makeatletter
\renewcommand\@dotsep{200}
\makeatother

\begin{document}
	{\hfill CERN-TH-2023-045}

\vspace{2cm}

\begin{center}
	\boldmath
	
	{\textbf{\LARGE  The Universal One-Loop Effective Action with Gravity}}
	
	\unboldmath
	
	\bigskip
	
	\vspace{0.5 truecm}
	
	{\bf R\'emy Larue}$^{a}$ and {\bf J\'er\'emie Quevillon}$^{a,b}$
	\\[5mm]
	{$^a$\it Laboratoire de Physique Subatomique et de Cosmologie,}\\
	{\it Universit\'{e} Grenoble-Alpes, CNRS/IN2P3, Grenoble INP, 38000 Grenoble, France}\\[2mm]
	{$^b$\it CERN, Theoretical Physics Department, Geneva 23 CH-1211, Switzerland}\\[2mm]

	\vspace{2cm}
	
	{\bf Abstract }
\end{center}

\begin{quote}

We complete the so-called Universal One-Loop Effective Action (UOLEA) with effects of gravity and provide a systematic approach to incorporate higher dimensional operators in curved spacetime. The functional determinant stemming from the path integral is computed using the Covariant Derivative Expansion (CDE), in a momentum representation that does not rely on a specific choice of coordinate to be defined, as it often is. This efficient approach manifests an interesting novelty as it allows to integrate out chiral fermions in curved spacetime in a direct manner leading to new operators involving the curvature,  and provides a new alternative to the use of Feynman diagrams in that regard. The method presented would very well fit in a code that performs CDE, offering the possibility to integrate out at one-loop fields on a curved spacetime background, including spin-2 fields, like the graviton. Eventually these results should provide an interesting way to study low energy effects of UV completions of gravity.
\end{quote}

\thispagestyle{empty}
\vfill

\newpage

\tableofcontents

\newpage

\section{Introduction}

The interest in Effective Field Theory (EFT) techniques have been gathering momentum recently due to the lack of new physics at the weak scale. It seems to indicate that the Standard Model (SM)  should be supplemented by higher dimensional operators and then considered itself as an EFT. The so-called matching between a UV theory and the effective theory at a given energy scale was in the first place performed using Feynman diagrams. However, the functional approach soon turned out to be quite elegant, powerful and optimal. One formally integrates out a heavy degree of freedom within the path integral to obtain directly the effective action at a given loop order. On top of being less computationally involved, the functional approach is more systematic. It allowed to unravel the universal structure of the one-loop effective action, which is expressed in the Universal One-Loop Effective Action (UOLEA)
 \cite{Henning:2014wua,Drozd:2015rsp,Ellis:2016enq,Ellis:2017jns,Ellis:2020ivx}.

At one-loop, the effective action stemming from the integration of the heavy fields arises in the form of a functional determinant. Different methods can be employed to decipher this functional determinant, such as the heat kernel based on position representation \cite{DeWitt:1964mxt,Schwinger:1951nm}, or more recently the Covariant Derivative Expansion (CDE) based on momentum representation \cite{Henning:2014wua,Drozd:2015rsp}. The latter was enhanced to encompass mixed heavy-light contributions \cite{Ellis:2016enq}, diagrammatic representation \cite{Zhang:2016pja}, integration of a heavy chiral fermion   (i.e involving $\gamma_5$ couplings)\cite{Ellis:2020ivx}, UV theory involving derivative couplings \cite{Quevillon:2021sfz}, and even the evaluation of generic QFT anomalies \cite{Filoche:2022dxl}. There are also prospects of beyond one-loop functional methods \cite{Alonso:2022ffe,vonGersdorff:2022kwj}.

Functional methods are well fit to comprise the effects of the curvature of spacetime. It was undertaken in both the heat kernel \cite{Schwinger:1951nm,DeWitt:1964mxt,DeWitt:1967yk,DeWitt:1967ub,Utiyama:1962sn,Fradkin:1976xa,Barvinsky:1985an,Vilkovisky:1992pb,Avramidi:2000bm} and the CDE \cite{Binetruy:1988nx,Alonso:2019mok}. As opposed to a Feynman diagram approach (see for example~\cite{Inagaki:1997kz,Buchbinder:1992rb}), gravity needs not be linearised to obtain the gravitational loop corrections. Even though they are more attainable, the higher order corrections remain a computational challenge to obtain.

The previous CDE procedures in curved spacetime relied on the use of the so-called Gaillard-Cheyette sandwish \cite{Cheyette:1987qz} to form covariant operators. Although it provides a manifestly covariant expansion, it also makes the computation much more intricate. In the CDE presented in this paper, this step is avoided. Together with the use of convenient choice of gauge and coordinate system, it makes the computation of higher order corrections in curved spacetime more systematic and thus easier to compute. For the first time, the non-renormalisable corrections are obtained within the framework of the CDE, and on a generic spacetime background.

Although a specific choice of coordinates can simplify the computation, our expansion is coordinate independent. Particularly, the question of the Fourier transform in curved spacetime is treated so as to obtain a diffeomorphism invariant expansion, whereas former approaches were mostly relying on a specific choice of coordinate, the Riemann Normal Coordinates (RNC), to define it \cite{Bunch:1979uk,Binetruy:1988nx,Parker:2009uva}. As a result, the method can also be used to obtain non-covariant results such as consistent gravitational anomalies.

Another novelty of this paper is the derivation of a fermionic CDE in curved spacetime. Previous methods (heat kernel, CDE and more recently using the worldline formalism \cite{Bastianelli:2008cu}) always relied on a generic bosonic form of the functional determinant. It can describe the effective action after integrating out real and complex scalar fields, massless and massive gauge bosons, the spin-2 metric field, and even vector-like fermions. Despite its generic form, it cannot describe a chiral fermion, as was pointed out in \cite{Ellis:2020ivx}. For the first time, a chiral fermion in curved spacetime is integrated out within the functional approach in a universal form, and leads to new renormalisable and non-renormalisable operators that were not computed before. It also provides a new alternative to the use of Feynman diagrams.  Besides the computational simplicity that is proper to the path integral approach with respect to the use of Feynman diagrams, it has the advantage of not needing to perturb the metric around a flat background, which significantly simplifies the calculations.  

Our result is the one-loop action in curved spacetime
up to six dimensional operators in the bosonic CDE, and up to five in the fermionic CDE,  given in a close form universal formula. Our systematic procedure can be used in practice to obtain much higher dimensional operators. This Gravitational version of the UOLEA should be significantly useful to study low energy consequences of the UV completion of gravity, or generical models including heavy degrees of freedom in gravity (see for example \cite{Nakonieczny:2018djb}).

The paper is organised as follows. In Section~\ref{sec2}, we acquaint the reader  with the CDE and the UOLEA in flat spacetime. In Section~\ref{sec3}, we introduce our notations and we derive the CDE in curved spacetime both bosonic and fermionic. We also provide in details the systematic procedure to perform the CDE. We compute the operators up to dimension 6 of the universal bosonic one-loop effective action in curved spacetime in Section~\ref{sec4}, and of up to dimension 5 of the universal fermionic one-loop effective
action in curved spacetime in Section~\ref{sec5}. In this last section we connect our results with an example of computation of the axial-gravitational anomaly. The Appendix~\ref{Appendix:master-integrals} outlines the computation of the associated master integrals, and in Appendix~\ref{Appendix:MomentumManifold} we give details about the coordinate independent Fourier transform on a generic manifold. The Appendix~\ref{Appendix:RNCFS} provides RNC and Fock-Schwinger (FS) gauge formulae.

\section{UOLEA in flat spacetime}
\label{sec2}
In this Section, we outline the UOLEA and CDE methods in flat spacetime.
We start from an action for the UV theory, $S[\phi,\Phi]$, that depends on light degrees of freedom $\phi$ and heavy ones $\Phi$. The effective action after integrating out the heavy degrees of freedom is,
\begin{equation}
iS_{\mathrm{eff}}[\phi]=\log \int \dc\Phi e^{iS[\phi,\Phi]}\, .
\end{equation}
To perform the path integral, we expand $\Phi$ around its background value $\Phi=\Phi_c+\eta$ such that $\frac{\delta S}{\delta \Phi}[\Phi_c]=0$,
\begin{align}
\begin{split}
iS_{\mathrm{eff}}[\phi]&=\log\int \dc\eta e^{iS[\phi,\Phi_c]+\frac{i}{2}\eta\cdot\frac{\delta^2 S}{\delta\Phi^2}[\Phi_c]\cdot\eta+\mathcal{O}(\eta^3)}\\
&\simeq iS[\phi,\Phi_c]-\frac{1}{2}\Tr\log\frac{\delta^2 S}{\delta\Phi^2}[\phi,\Phi_c]\, ,
\end{split}
\end{align}
where the functional trace \textit{Tr} is both over spacetime and internal indices.

There are several methods in the literature to expand the functional trace, but the one we will employ is the covariant derivative expansion (CDE) \cite{Gaillard:1985uh,Cheyette:1987qz,Henning:2014wua}. We can then assume a general form for the second derivative of the action to derive the so-called Universal One-Loop Effective Action (UOLEA) \cite{Drozd:2015rsp,Ellis:2016enq,Ellis:2017jns,Ellis:2020ivx}.
The CDE can also be performed when the mass matrix is non-degenerate \cite{Drozd:2015rsp}, to encompass mixed heavy-light loops \cite{Ellis:2016enq,Ellis:2017jns}, to integrate out massive chiral fermions \cite{Ellis:2020ivx,Quevillon:2021sfz}  and a diagrammatic approach also exists to help with the expansion \cite{Zhang:2016pja}.

\subsection{Bosonic UOLEA}
\label{BosUOLEAflat}

We assume the following form for the second derivative of the action of the UV theory,
\begin{equation}
S^{\mathrm{eff}}_{\mathrm{1 loop}}=ic_s\Tr\log\left( D^2 + m^2 +U \right)\, ,
\label{S1loopbos}
\end{equation}
where $D$ is the covariant derivative that bears the gauge connections, and $U$ is some local operator (it bears no open derivative). The constant $c_s$ depends on the heavy field that is integrated out. If it is a real scalar, complex scalar, vector-like fermion, gauge boson or Fadeev-Popov ghost then it takes the value $1/2$, $1$, $-1/2$, $1/2$ and $-1$ respectively. Note that the explicit form of $U$ also depends on the nature of the heavy field.

We evaluate the trace over spacetime by inserting a complete set of spatial and momentum eigenstates,
\begin{align}
\begin{split}
S^{\mathrm{eff}}_{\mathrm{1 loop}}&=ic_s\int \dd^d x\frac{\dd^d q}{(2\pi)^d} e^{-iq\cdot x}\tr\log\left( D^2+m^2+U \right)e^{iq\cdot x}\\
&=ic_s\int \dd^d x\frac{\dd^d q}{(2\pi)^d} \tr\log\left( (D+iq)^2+m^2+U \right)\\
&=ic_s\int \dd^d x\frac{\dd^d q}{(2\pi)^d} \tr\log\Big( -\Delta^{-1}\big( 1-\Delta\left( D^2+2iq\cdot D+U \right)\big)\Big)\, ,
\end{split}
\end{align}
where $\Delta=1/(q^2-m^2)$ and the remaining trace is over internal indices (gauge, spin, \dots).

An extra step introduced in \cite{Gaillard:1985uh,Cheyette:1987qz} can be undertaken. It consists in sandwishing,\newline \noindent $e^{-iq\cdot x}\tr\log\left( D^2+m^2+U \right)e^{iq\cdot x}$, with $e^{\pm iD\cdot\pa_q}$ which has the advantage of stowing the covariant derivatives inside commutators, but at the cost of rendering the computation more tedious. Besides, because $\pa_q \equiv \frac{\partial}{\partial q}$ does not commute with $\Delta$, the logarithm cannot be expanded simply. One has to write the logarithm as the primitive of an inverse function and expand the inverse, as we will do in curved spacetime \footnote{It is also possible to use the Baker-Campbell-Hausdorff (BCH) formula to expand the logarithm.}.

Since $D^2+2iq\cdot D +U$ and $\Delta$ commute, we can expand the logarithm,
\begin{equation}
S^{\mathrm{eff}}_{\mathrm{1 loop}}=-ic_s\int \dd^d x\frac{\dd^d q}{(2\pi)^d} \sum_{n\geq1}\frac{1}{n}\Big[\Delta\left(D^2+2iq\cdot D+U\right)\Big]^n\, .
\end{equation}

Conveniently, a factorisation between the momentum integrals and the operator part occurs which is the origin of the computation of the UOLEA \cite{Henning:2014wua,Drozd:2015rsp}, and allows to derive the Wilson coefficients in terms of master integrals (App.~\ref{Appendix:master-integrals}). The UOLEA operator structures, written in terms of the matrices D and U, become EFT operators when substituting in the specific forms of these matrices (in terms of the light fields and for a given UV model) which can then be cast into the desired non-redundant EFT basis. In light of its generality, this suggests that in future calculations of one-loop Wilson coefficients for operators of dimension higher than four (in practice up to six dimensions) one can skip the usual Feynman diagram or path-integral methods and proceed directly to the UOLEA master equation as the starting point.

\subsection{Fermionic UOLEA}
\label{UOLEAferm}

If we integrate out a massive chiral fermion~\footnote{
Let us make a comment on massive chiral fermions. The mass term is a hard breaking source of axial symmetries (local or global). These symmetries can be made manifest at tree-level by implementing their spontaneous breaking and introducing their associated Goldstone bosons. In this paper, we choose for convenience to work within the unitary basis and loose manifest tree-level axial invariance.
} which UV action is,
\begin{equation}
S=\int\dd^dx\bar\psi\left(i\slashed D-m-Q\right)\psi\, ,
\end{equation}
we obtain a one-loop effective action of the form,
\begin{equation}
S^{\mathrm{eff}}_{\mathrm{1 loop}}=-i\log\det\left( i\sD-m-Q \right)\, .
\label{AnsatzFerm}
\end{equation}

This determinant can be bosonised by squaring it~\footnote{ Note that it can also be bosonised by multiplying by the hermitian conjugate, which amounts to computing the modulus of the determinant. Compared to the square of the determinant, only a phase is omitted. This phase is relevant for computing consistent anomalies~\cite{Alvarez-Gaume:1983ihn,Filoche:2022dxl}.}. We split $Q$ into $Q_e$ which has an even number of Dirac matrices and $Q_o$ with an odd number,
\begin{align}
\begin{split}
\log\det\left(i\sD-m-Q\right)&=\frac{1}{2}\log\det\left(i\sD-m-Q_e-Q_o\right)\det\left(-i\sD-m-Q_e+Q_o\right)\\
&=\frac{1}{2}\log\det\Big(D^2+m^2+\frac{1}{4}[\gamma^\mu,\gamma^\nu]F_{\mu\nu}+2mQ_e\\
&\hspace{2,5cm}+Q(Q_e-Q_o)-[i\sD,Q_e]+\{i\sD,Q_o\}\Big)\, ,
\label{AnsatzBos2}
\end{split}
\end{align}
where in the first line we used the vanishing of the trace of an odd number of Dirac matrices, therefore the invariance under flipping their sign.

Eq.~\eqref{AnsatzBos2} reduces to a determinant of the form Eq.~\eqref{S1loopbos} only if $Q_o=0$, with $U=\frac{1}{4}[\gamma^\mu,\gamma^\nu]F_{\mu\nu}+2mQ_e+Q_e^2-[i\sD,Q_e]$. However, the usual ansatz used in the heat kernel method,  previous CDE approaches~\cite{Binetruy:1988nx,Alonso:2019mok} and the worldline formalism~\cite{Bastianelli:2008cu} assumes that $e^{-iq\cdot x} U e^{iq\cdot x}=U$ (i.e no open derivative), which is not always true depending on $Q_e$ (for example if $Q_e\supset \gamma_5$).

In other words, the usual ansatz Eq.~\eqref{S1loopbos} (used in heat kernel, CDE, worldline methods) does not encompass chiral fermions since it is not equivalent to Eq.~\eqref{AnsatzBos2}. Eq.~\eqref{AnsatzBos2} is cumbersome to work with, instead we choose to directly use the CDE on the fermionic determinant Eq.~\eqref{AnsatzFerm},
\begin{align}
\begin{split}
S^{\mathrm{eff}}_{\mathrm{1 loop}}&=-i\int\dd^dx\frac{\dd^dq}{(2\pi)^d}\tr\,\log\Big( \Delta^{-1}\left( 1-\Delta\left(-i\sD+Q\right) \right) \Big)\\
&=-i\int\dd^dx\frac{\dd^dq}{(2\pi)^d}\tr\,\sum_{n\geq1}\frac{1}{n}\Big[\Delta\left(-i\sD+Q\right)\Big]^n\, ,
\end{split}
\end{align}
where now $\Delta=m/(q^2-m^2)-\slashed q/(q^2-m^2)$\, .

If we assume a general form for $Q$, namely a scalar $W_0$, pseudo-scalar $W_0\gamma_5$, vector $\gamma^\mu V_\mu$ and pseudo-vector part $\gamma^\mu A_\mu\gamma_5$, we can derive the so-called fermionic UOLEA \cite{Ellis:2020ivx}. The case of UV theories involving derivative couplings (such as axion models) requires extra care and is treated in details in ~\cite{Quevillon:2021sfz}.\\

\section{Curved spacetime CDE}
\label{sec3}

We now turn to the presentation of the CDE in curved spacetime.
In the paper we adopt standard conventions regarding gravity such as assuming the Levi-Civita connection. $\nabla$ is the covariant derivative with the Christoffel connection only. We use the following conventions,
\begin{align}
\begin{split}
\nabla_\mu v_\nu=\pa_\mu v_\nu-\Gamma^\lambda_{\mu\nu}v_\lambda\text{,} \quad \nabla_\mu v^\nu=\pa_\mu v^\nu+\Gamma^\nu_{\mu\lambda}v^\lambda\, .
\end{split}
\end{align}
The Riemann tensor and Ricci tensor are defined by,
\begin{equation}
 [\nabla_\alpha,\nabla_\beta]v^\mu=\tensor{R}{^\mu_\rho_\alpha_\beta}v^\rho\, ,\quad\quad R_{\mu\nu}=\tensor{R}{^\lambda_\mu_\lambda_\nu} \ . 
\label{RiemannConvention}   
\end{equation}

We define the general covariant derivative $\dc$ and field strength $\mathcal{F_{\mu\nu}}=[\dc_\mu,\dc_\nu]$ which include all the connections (Christoffel, gauge, spin-connection). Whereas the covariant derivative and field strength that bear only the gauge and spin-connection are denoted by $D$ and $F$ such that for a field $\Phi$,
\begin{equation}
\dc_\mu\Phi=D_\mu\Phi\, ,\quad \mathcal{F}_{\mu\nu}\Phi=F_{\mu\nu}\Phi\, .
\end{equation}

A scalar field behaves trivially when the curvature of spacetime is introduced. For a scalar field charged under a gauge group associated to the gauge field $V$, we have,
\begin{equation}
D_\mu=\pa_\mu+iV_\mu,\quad F_{\mu\nu}=i(\pa_\mu V_\nu)-i(\pa_\nu V_\mu)-[V_\mu,V_\nu]\, .
\end{equation}

\noindent The covariant derivative acting on a fermion is,
\begin{equation}
D_\mu=\partial_\mu+iV_\mu+\omega_\mu \, ,
\end{equation}
where the spin-connection is $\omega_\mu=\frac{1}{8}[\gamma^a,\gamma_b] \tensor{e}{_a^\nu}(\pa_\mu \tensor{e}{^b_\nu}-\Gamma^\lambda_{\mu\nu}\tensor{e}{^b_\lambda})$, with $\tensor{e}{^a_\mu}$ the orthonormal tangent frame vielbein such that $g_{\mu\nu}=\tensor{e}{^a_\mu} \tensor{e}{^b_\nu} \eta_{ab}$ (latin indices referring to the tangent frame). We follow \cite{Parker:2009uva} for conventions for the spin-connection. As a consequence, the fermion field strength is,
\begin{equation}
F_{\mu\nu}=\frac{1}{4}\gamma^\rho\gamma^\sigma R_{\mu\nu\rho\sigma} +i(\pa_\mu V_\nu)-i(\pa_\nu V_\mu)-[V_\mu,V_\nu]\, ,
\label{eq:Fgrav}
\end{equation}
and,
\begin{align}
\begin{split}
&\slashed \dc^2=\dc^2-\frac{i}{2}\sigma^{\mu\nu}\mathcal{F}_{\mu\nu} \text{ where } \sigma^{\mu\nu}=\frac{i}{2}[\gamma^\mu,\gamma^\nu]\\
&\frac{i}{2}\sigma.F\,=\frac{1}{4}R\, \mathds{1}_{\mathrm{Dirac}}\, ,
\label{eq:sigmaFgrav}
\end{split}
\end{align}
where $\mathds{1}_{\mathrm{Dirac}}$ is the identity in Dirac space.

The fermion can also couple to an axial gauge field $A_\mu$. However, it proves convenient in the computation to separate it from the covariant derivative as we will do in Section~\ref{CDEfermion}.

In curved spacetime, the Dirac matrices are defined with respect to the Dirac matrices in the tangent frame, and we have the Clifford algebra,
\begin{equation}
\gamma^\mu=\tensor{e}{_a^\mu}\gamma^a\, ,\quad \{\gamma^\mu,\gamma^\nu\}=2g^{\mu\nu}\, ,
\end{equation}
that we used to derive Eq.~\eqref{eq:sigmaFgrav}.

Finally, the general covariant derivative commutes with the Dirac matrices,
\begin{align}
\label{nablagamma}
(\dc_\mu\gamma^\nu)=(\pa_\mu\gamma^\nu)+\Gamma^\nu_{\mu\rho}\gamma^\rho+[\omega_\mu,\gamma^\nu]=0\, ,\quad
(\dc_\mu \gamma_5)=[\omega_\mu,\gamma_5]=0 \, .
\end{align}

\subsection{Fourier transform in curved spacetime}
\label{MomRep}

Previous litterature \cite{Bunch:1979uk,Binetruy:1988nx,Parker:2009uva} involving momentum representation in curved spacetime relied on a specific choice of coordinate, the RNC where spacetime is locally flat around a point, to define the Fourier transform. However, we would like to define the Fourier transform without relying on a specific choice of coordinate (which is of relevance when dealing with quantum anomalies for example). It does not seem trivial to us that the usual Fourier transform in curved spacetime leads to a coordinate independent result, since the choice of momentum representation depends on the choice of coordinate. We explain our procedure for defining the Fourier transform in curved spacetime, and show that it is indeed independent of the choice of coordinate

A manifold $\mathcal{M}$ of dimension $d$ is locally mapped to $\mathds{R}^d$ by a coordinate system. On a given subset of $\mathcal{M}$, with coordinate $x^\mu$, we can use the momentum $q_\mu$ associated to $x^\mu$ such that $\pa q_\nu/\pa x^\nu=0$, as we would do in the flat space $\mathds{R}^d$.

The issue is that $x\cdot q$ is not invariant under diffeomorphism, because $x^\mu \pa_\mu$ does not transform as a vector. Nonetheless, we show in Appendix~\ref{Appendix:MomentumManifold} that $e^{-iq\cdot x}\tr\,\mathcal{O}(x(p),i\pa_{x})e^{iq\cdot x}$ as well as the measure $\dd^d x\, \dd^d q/(2\pi)^d$ are diffeomorphism invariant, provided the operator $\mathcal{O}$ is covariant ~\footnote{If $\mathcal{O}$ is not covariant, the expansion provides the correct non-covariant quantity as opposed to an expansion that would rely on a specific choice of coordinate.}.

This allows us to write the functional trace in a diffeomorphism invariant manner using the momentum representation,
\begin{equation}
\Tr\,\mathcal{O}=\int_{p\in\mathcal{M}}\dd^d x(p)\frac{\dd^d q}{(2\pi)^d}e^{-iq\cdot x(p)}\tr\,\mathcal{O}(x(p),i\pa_{x})e^{iq\cdot x(p)}=\int_{p\in\mathcal{M}}\dd^d x(p)\frac{\dd^d q}{(2\pi)^d}\tr\,\mathcal{O}(x(p),i\pa_{x}-q)\, .
\end{equation}
Throughout the expansion of the functional trace, we will make use of,
\begin{equation}
\frac{\pa q_\mu}{\pa x^\nu}=0\, ,\quad (\pa_\mu q^2)=(\pa_\mu g^{\alpha\beta})q_\alpha q_\beta\, ,
\end{equation}
to commute the momentum dependence to left of the covariant derivatives.

\subsection{Bosonic CDE in curved spacetime}
We seek to compute a functional trace of the form,
\begin{equation}
    S^{\mathrm{boson}}_{\mathrm{eff}}=ic_s\,\Tr \log\, \left(\g(g^{\mu\nu}\dc_\mu \dc_\nu+m^2+U)\right)\, ,
\end{equation}
where $\dc_\mu$ is again the general covariant derivative, and $U$ is some local operator (i.e it doesn't act on everything to its right, as opposed to an open derivative). The trace above is both over internal spaces and spacetime. $c_s$ depends on the nature of the field that is integrated out ( see Sec.~\ref{BosUOLEAflat}). As explained in Sec.~\ref{MomRep}, the functional trace in curved spacetime is written in a diffeomorphism invariant manner as,
\begin{align}
\begin{split}
S^{\mathrm{boson}}_{\mathrm{eff}}=ic_s\int \dd^dx \frac{\dd^d q}{(2\pi)^d}e^{-iq\cdot x}\tr\log\,\g(\dc^2+m^2+U)e^{iq\cdot x}\, . 
\end{split}
\end{align}
Similarly as in the flat spacetime case, we introduce the propagator $\Delta=1/(q^2-m^2)$,
\begin{equation}
S^{\mathrm{boson}}_{\mathrm{eff}}=ic_s\int \dd^dx \frac{\dd^d q}{(2\pi)^d}\tr\log\,\g\left(-\Delta^{-1}(1-\Delta(\dc^2+iq\cdot \dc+\dc\cdot iq+U)\right)\, ,
\end{equation}
In curved spacetime one significant novelty and difficulty come from the fact that $\Delta$ and $\dc$ do not commute anymore since,
\begin{equation}
[\dc_\mu,\Delta]=-(\pa_\mu q^2)\Delta^2\, .
\end{equation}
Therefore one cannot expand the \textit{log} directly as in flat spacetime. We thus rely on the following trick to perform the expansion \footnote{It is also possible to expand directly the \textit{log} of non-commuting operators using the Baker-Campbell-Hausdorf formula.}: we rewrite the \textit{log} as the primitive of the inverse function and then expand it using,
\begin{equation}
    \frac{1}{A^{-1}(1-AB)}=\sum_{n\geq 0}(AB)^n A\, ,
\end{equation}
which does not require the matrices $A$ and $B$ to commute.

To make an inverse function appear we write (see \cite{Henning:2014wua} for example),
\begin{equation}
\log \g(\dc^2+m^2+U(m))=\int^{m^2} \dd m'^2\frac{1}{(\dc^2+m'^2+U(m))}\, .
\end{equation}
We thus obtain,
\begin{align}
\begin{split}
S^{\mathrm{boson}}_{\mathrm{eff}}=&ic_s\int \dd^dx \frac{\dd^d q}{(2\pi)^d}\int^{m^2} \dd m'^2\tr\,\frac{1}{-\Delta^{-1}(1-\Delta(\dc^2+iq\cdot \dc+\dc\cdot iq+U)}\\
=&-ic_s\int \dd^dx \frac{\dd^d q}{(2\pi)^d}\int^{m^2} \dd m'^2\tr\,\sum_{n\geq 0}\left[\Delta(\dc^2+2iq\cdot \dc-g^{\mu\nu}\Gamma^\rho_{\mu\nu}q_\rho+U)\right]^n\Delta\, ,
\label{MasterFormula1}
\end{split}
\end{align}
where now $\Delta=1/(q^2-m'^2)$. 
Note that the spacetime measure $\g$ disappears in the expansion. If the masses are non-degenerate, we can multiply the mass matrix by a parameter and integrate over this parameter instead (see \cite{Drozd:2015rsp} for example).

Note that when all the Lorentz indices to the right of a covariant derivative $\mathcal{D}$ are contracted among themselves, the Christoffel connection cancels (e.g $(\mathcal{D}_\mu v^\nu u_\nu)=(\pa_\mu v^\nu u_\nu)$). In Eq.~\eqref{MasterFormula1} the only covariant derivative that has uncontracted indices to its right is $\mathcal{D}_\mu$ in $\mathcal{D}^2=g^{\mu\nu}\mathcal{D}_\mu\mathcal{D}_\nu$. Acting on a field $\phi$ we have,
\begin{equation}
\mathcal{D}^2\phi=g^{\mu\nu}\mathcal{D}_\mu D_\nu\phi=g^{\mu\nu}D_\mu D_\nu\phi-g^{\mu\nu}\Gamma^\rho_{\mu\nu}D_\rho\phi\, .
\end{equation}
Once $\mathcal{D}^2$ is written as such, all the $\mathcal{D}$ have only contracted indices to their right, hence they can be replaced by $D$,
\begin{align}
\begin{split}
S^{\mathrm{boson}}_{\mathrm{eff}}=-ic_s\int \dd^dx \frac{\dd^d q}{(2\pi)^d}\int \dd m'^2\tr\,\sum_{n\geq 0}\left[\Delta(g^{\mu\nu}D_{\mu}D_\nu-g^{\mu\nu}\Gamma^\rho_{\mu\nu}D_\rho+2iq\cdot D-ig^{\mu\nu}\Gamma^\rho_{\mu\nu}q_\rho+U)\right]^n\Delta\, ,
\label{MasterFormula2}
\end{split}
\end{align}

Both Eq.~\eqref{MasterFormula1} and Eq.~\eqref{MasterFormula2} can be used for the expansion. In the former, $\dc$ contracts the Lorentz indices yielding Christoffel connections, but commutes with the metric. In the latter, $D$ does not contract the Lorentz indices but does not commute with the metric (which in the end yields Christoffel connections in virtue of $(\pa_\mu g^{\alpha\beta})=-\Gamma^\alpha_{\mu\lambda}g^{\lambda\beta}-\Gamma^\beta_{\mu\lambda}g^{\alpha\lambda}$).

Note that both in Eq.~\eqref{MasterFormula1} and Eq~\eqref{MasterFormula2}, if $U\supset [\gamma^\mu,\gamma^\nu]\mathcal{F}_{\mu\nu}$ from the bosonisation of a vector-like fermion, it is possible to rewrite it as $[\gamma^\mu,\gamma^\nu]F_{\mu\nu}$ since all the indices to the right of $\mathcal{F}_{\mu\nu}$ are contracted.

The master integrals produced by the bosonic expansion are of the form,
\begin{equation}
\int\dq q_{\mu_1}\dots q_{\mu_{2l}}\int^{m^2}\dd m'^2\,\frac{1}{(q^2-m'^2)^n}=\g\mathcal{J}[q^{2l}]^n\,g_{\mu_1\dots\mu_{2l}}\, ,
\end{equation}
where $g_{\mu_1\dots\mu_{2l}}$ is the fully symmetrised metric. They are related to the usual master integrals in flat spacetime $\mathcal{I}$ (see App.~\ref{Appendix:master-integrals}).

\subsection{Fermionic CDE in curved spacetime}
\label{CDEfermion}

The covariant derivative expansion can also be performed to expand a functional determinant of the form,
\begin{equation}
S^{\mathrm{fermion}}_{\mathrm{eff}}=-i\log\Tr\left(\g(i\sdc - m - Q)\right)\, .
\end{equation}
The fermion may be chiral. In that case, it is convenient to put the axial field in $Q\supset \slashed A\gamma_5$ and keep $(\dc_\mu\psi)=(\pa_\mu+iV_\mu+\omega_\mu)\psi$.

The functional trace is expressed as in the bosonic CDE and leads to,
\begin{equation}
S^{\mathrm{fermion}}_{\mathrm{eff}} =-i \int\dd^dx\frac{\dd^dq}{(2\pi)^d}\tr\log\g\left( i\sdc-\slashed q-m-Q \right)\, .
\end{equation}
We make the inverse function appear by integrating over the mass instead of integrating over the mass squared as previously~\footnote{Since all Lorentz indices are contracted, we could extract the Christoffel connection as previously and perform the expansion with $D$ instead of $\dc$ after writing either $\sdc=\gamma^\mu D_\mu$ or $\sdc=D_\mu\gamma^\mu+\Gamma^\mu_{\mu\nu}\gamma^\nu$. However it does not simplify the expansion since $[D_\mu,\gamma^\nu]\neq0$.},
\begin{align}
\begin{split}
S^{\mathrm{fermion}}_{\mathrm{eff}}=& i\int\dd^dx\frac{\dd^dq}{(2\pi)^d}\int^m \dd m'\,\tr\,\frac{1}{i\sdc-\slashed q-m'-Q(m)}\\
=&i\int\dd^dx\frac{\dd^dq}{(2\pi)^d}\int^m \dd m'\,\tr\,\sum_{n\geq0}\left[\Delta\left( -i\sdc+Q \right)\right]^n\Delta\, ,
\label{MasterFormulaFerm}
\end{split}
\end{align}
where now $\Delta=-1/(\slashed q + m')\,$, which can be split as,
\begin{equation}
\Delta=\frac{m'}{q^2-m'^2}+\frac{-\slashed q}{q^2-m'^2}\, .
\label{DeltaFermSplit}
\end{equation}
Again if the mass matrix is non-degenerate one just has to multiply the mass matrix by a parameter and integrate over this parameter instead.

This expansion will produce master integrals of the form,
\begin{equation}
\int\frac{\dd^d q}{(2\pi)^d} q_{\mu_1}\dots q_{\mu_{2l}} \int^m\dd m' \frac{m'^{k}}{(q^2-m'^2)^n}=\g\mathcal{K}[q^{2l}]^{k}_{n}\,g_{\mu_1\dots\mu_{2l}}\, ,
\end{equation}
with $n\geq k$. $g_{\mu_1\dots\mu_{2l}}$ is the fully symmetrised metric. They are related to the usual master integrals in flat spacetime $\mathcal{I}$ (see App.~\ref{Appendix:master-integrals}).

As a remark, using the integration over the mass, it is also possible to use the Gaillard and Cheyette sandwish mentionned in Section \ref{BosUOLEAflat} for the fermionic expansion without relying on the BCH formula to expand the logarithm.

\subsection{A systematic procedure}
\label{SystematicProcedure}

In flat spacetime, the factorisation of the momentum integrals from the operator part is key in deriving a universal formula. In curved spacetime, the momentum dependence does not commute anymore with the covariant derivatives. Nevertheless, we can recover the factorisation of the momentum integration after commuting carefully the momentum part through the covariant derivatives. Using $(\pa_\mu q_\nu)=0$, we derive a set of useful commutation relations presented in Tab.~\ref{tab1}.

\begin{table}[h!]
\begin{center}
\begin{tabular}{ |m{6cm}|m{7cm}| } 
  \hline
  General covariant derivative: $\dc$ & Gauge (and spin-connection) covariant derivative: $D$ \\ 
  \hline
    $$[\dc_\mu, q_\nu]= -\Gamma^\rho_{\mu\nu}q_\rho$$ $$[\dc_\mu, q^\nu]= -g^{\nu\sigma}\Gamma^\rho_{\mu\sigma}q_\rho$$ 
    $$[\dc_\mu,\gamma^\nu]=0$$ $$[\dc_\mu,g^{\nu\rho}]=0$$ $$[\dc_\mu,\Gamma^\nu_{\rho\sigma}]=(\nabla_\mu\Gamma^\nu_{\rho\sigma})$$ & $$[D_\mu, q_\nu]= 0$$ $$[D_\mu,q^\nu]=(\pa_\mu g^{\rho\nu})q_\rho$$ $$[D_\mu,\gamma^\nu]=(\pa_\mu\gamma^\nu)+[\omega_\mu,\gamma^\nu]=-\Gamma^\nu_{\mu\rho}\gamma^\rho$$ $$[D_\mu,g^{\nu\rho}]=(\pa_\mu g^{\nu\rho})$$ $$[D_\mu,\Gamma^\nu_{\rho\sigma}]=(\pa_\mu\Gamma^\nu_{\rho\sigma})$$ \\
    \hline
    \multicolumn{2}{|c|}{}\\
    \multicolumn{2}{|c|}{
    $[D_\mu,\delta]=[\dc_\mu,\delta]=(\pa_\mu\delta)=-(\pa_\mu q^2)\delta^2=-(\pa_\mu g^{\alpha\beta})q_\alpha q_\beta\delta^2$
    }\\
    \multicolumn{2}{|c|}{}\\
    \multicolumn{2}{|c|}{
    $[\dc_\mu,-\slashed q \delta]=-\gamma^\alpha \left( (\nabla_\mu q_\alpha)\delta + q_\alpha(\pa_\mu\delta) \right)=-\gamma^\alpha\left( -\Gamma^\lambda_{\mu\alpha}q_\lambda - q_\alpha (\pa_\mu g^{\rho\sigma})q_\rho q_\sigma \delta^2  \right)$
    }\\
    \multicolumn{2}{|c|}{}\\
    \hline
\end{tabular}
\caption{Set of commutation rules. With the notation $\delta=1/(q^2-m^2)$.}
\label{tab1}
\end{center}
\end{table}

Once the momentum dependence is commuted to the left of the covariant derivatives, the integration over momentum and mass can be performed \footnote{The mass integration variable commutes with every operators, so no difficulty arises in that regard.}. Then the different terms have to be combined together to form covariant quantities. This last point may seem a tedious task since our expansion is not manifestly covariant, but we will see in the examples that the use of Riemann Normal Coordinates (RNC) and Fock-Schwinger (FS) gauge effortlessly provide the result in terms of covariant quantities.
Besides, the use of RNC from the beginning of the computation greatly reduces the number of terms to compute, and simplifies the commutations through the covariant derivatives.

In the following, we give examples to illustrate the systematic procedure to perform the CDE in curved spacetime: commute the momentum dependence to the left using Table~\ref{tab1}, perform the mass and momentum integration, form covariant quantities.
We will show that the use of RNC is not only useful to form covariant curvature quantities, but also from the first step of the procedure it reduces the number of terms that contribute at a given order, and simplifies the commutation of the momentum.

\subsubsection*{Example of computation - Bosonic CDE}

We will compute the first order term ($m^2$) of the bosonic UOLEA in curved spacetime to illustrate the procedure.

We will use the expansion from Eq.~\eqref{MasterFormula2}. For simplicity, we denote $D^2=g^{\mu\nu}D_\mu D_\nu$, $\Gamma D=g^{\mu\nu}\Gamma^\rho_{\mu\nu}D_\rho$, and $\Gamma q=g^{\mu\nu}\Gamma^\rho_{\mu\nu}q_\rho$, and we take $U=0$. The contributions at order $m^2$ are,
\begin{align}
\begin{split}
\left.\Lagr^{\mathrm{bos}}_{\mathrm{eff}}\right|_{\mathcal{O}(m^2)}=-ic_s\int\frac{\dd^dq}{(2\pi)^d}\int^{m^2}\dd m'^2\,\tr\,\Big( \Delta D^2\Delta + \Delta (2iq\cdot D-\Gamma D-i\Gamma q) \Delta (2iq\cdot D-\Gamma D-i\Gamma q) \Delta\Big)\, .
\label{m2Example1}
\end{split}
\end{align}

The first step is to commute the momentum dependence to the left, that is to say, commute the covariant derivatives to the right. For example, consider the first term of Eq.~\eqref{m2Example1},
\begin{equation}
\tr\,\Delta D^2\Delta=\tr\,\left(\Delta (\Delta D^2+(D^2\Delta)+2g^{\mu\nu}(D_\mu\Delta)D_\nu)\right)\, .
\label{D2DeltaExample}
\end{equation}
We then use the commutation relations from Table~\ref{tab1},
\begin{align}
&\tr\, (D^2\Delta)=\tr\,g^{\mu\nu} (\pa_{\mu\nu}\Delta) =\tr\,g^{\mu\nu}\left( -(\pa_{\mu\nu} q^2)\Delta^2+2(\pa_\mu q^2)(\pa_\nu q^2)\Delta^3 \right)\nonumber\\
&\tr\, 2g^{\mu\nu}(D_\mu\Delta)D_\nu=-2g^{\mu\nu}\tr\,(\pa_\mu q^2)\Delta^2 D_\nu\, .
\end{align}
Now we can perform the integration over mass and momentum and express it in terms of master integrals (App.~\ref{Appendix:master-integrals}).

The same procedure has to be applied for the 9 other terms from Eq.~\eqref{m2Example1}. Once the master integrals are explicited, all the terms can be combined together to form covariant quantities.\\

As emphasised earlier, since our expansion is diffeomorphism independent we can choose a specific coordinate system to simplify the computation. We will use the Riemann Normal Coordinates (RNC) around a point $x_0$. At $x=x_0+y$, the metric and the Christoffel symbols can be expanded around as,
\begin{align}
\begin{split}
&g_{\mu\nu}(y)=\eta_{\mu\nu}-\frac{1}{3}\tensor{R}{_\mu_\alpha_\nu_\beta}(x_0)y^\alpha y^\beta +\mathcal{O}(y^3)\\
&\Gamma^\mu_{\nu\rho}(y)=0-\frac{1}{3}(\tensor{R}{^\mu_\nu_\rho_a}+\tensor{R}{^\mu_\rho_\nu_a})(x_0)y^a+\mathcal{O}(y^2)\, .
\end{split}
\end{align}
We can use the RNC to help form covariant quantities after the expansion performed above, but simplifications occur starting from Eq.~\eqref{m2Example1}. The Christoffel symbols vanish at $x_0$, only their derivatives survive. We can already rule out from Eq.~\eqref{m2Example1} the terms that have Christoffel symbols without derivative to their left,
\begin{align}
\begin{split}
\left.\Lagr^{\mathrm{bos}}_{\mathrm{eff}}\right|_{\mathcal{O}(m^2)}=-ic_s\int\frac{\dd^dq}{(2\pi)^d}\int^{m^2}\dd m'^2\,\tr\,\Big( \Delta D^2\Delta + \Delta 2iq\cdot D \Delta (2iq\cdot D-\Gamma D-i\Gamma q) \Delta\Big)\, .
\label{m2Example2}
\end{split}
\end{align}
We then commute the momentum dependence to the left in RNC. For the term $\Delta D^2\Delta$, we obtained Eq.~\eqref{D2DeltaExample}. But the term including $(D_\mu\Delta)$ is proportional to a first derivative of the metric hence it vanishes. We are left with,
\begin{align}
\begin{split}
\int\frac{\dd^d q}{(2\pi)^d}\int^{m^2}\dd m'\,\tr\,\Delta D^2\Delta&=\int\frac{\dd^d q}{(2\pi)^d}\int^{m^2}\dd m'\tr\,\left(\Delta^2 D^2 +(D^2\Delta) \right)\\
&=\int\frac{\dd^d q}{(2\pi)^d}\int^{m^2}\dd m'\tr\,\left(\Delta^2 D^2 - \Delta^2 q_\alpha q_\beta g^{\mu\nu}(\pa_{\mu\nu} g^{\alpha\beta})\right)\\
&=\g\tr\,\left(\mathcal{J}[q^0]^2 D^2 - \mathcal{J}[q^2]^3 \frac{2}{3}R \right)\, .
\label{m2term1}
\end{split}
\end{align}
The second term from Eq.~\eqref{m2Example2} is also rather simple. Using Table~\ref{tab1} in RNC it reduces to,
\begin{align}
\begin{split}
&\int\frac{\dd^d q}{(2\pi)^d}\int^{m^2}\dd m'\tr\,\Delta 2iq\cdot D \Delta (2iq\cdot D-\Gamma D-i\Gamma q)\\
&=\int\frac{\dd^d q}{(2\pi)^d}\int^{m^2}\dd m'\tr\,\left( \Delta^2 (2i)^2 q^\mu q^\nu D_\mu D_\nu - 2iq^\mu g^{\alpha\beta}(\pa_\mu\Gamma^\rho_{\alpha\beta})D_\rho +2q^\mu g^{\alpha\beta}(\pa_\mu\Gamma^\rho_{\alpha\beta})q_\rho \right)\\
&=\g \mathcal{J}[q^2]^3 \tr\,\left( \frac{4}{3}R-4 D^2 \right)\, .
\label{m2term2}
\end{split}
\end{align}
Note that the integration with an odd power in $q$ in the numerator vanishes.

The next step which is to combine the different terms to form covariant quantities is avoided as far as the Christoffel part is concerned since the RNC provide directly the covariant quantities. There remains to form covariant quantities with the covariant derivatives, which can also by simplified using the Fock-Schwinger (FS) gauge (see details in App.~\ref{Appendix:RNCFS}).

Combining Eqs.~\eqref{m2term1} and \eqref{m2term2}, we obtain the one-loop effective action at order $m^2$,
\begin{equation}
\left.\Lagr^{\mathrm{bos}}_{\mathrm{eff}}\right|_{\mathcal{O}(m^2)}=\g\frac{c_s}{16\pi^2}m^2\left(1-\lm\right)\tr\,\frac{R}{6}\, .
\end{equation}
The remaining trace is over gauge and spin degrees of freedom. $\mu$ is the renormalisation scale from dimensional regularisation. We used the $\overline{MS}$ scheme, and will do so throughout this paper.

\subsubsection*{Example of computation - Fermionic CDE}

For completion, we briefly outline the computation of the $m^2$ term in the fermionic expansion, although it is similar to the procedure of the bosonic CDE. We take $Q=0$ and omitt the gauge sector for simplicity, the contribution at this order is,
\begin{equation}
\left.\Lagr^{\mathrm{ferm}}_{\mathrm{eff}}\right|_{\mathcal{O}(m^2)}=-i\int\frac{\dd^d q}{(2\pi)^d}\int^m\dd m'\,\tr\,\Delta\sdc\Delta\sdc\Delta\, .
\end{equation}
We first commute the momentum dependence to the left, we can do it in RNC to keep it simple. Let's keep in mind that the propagators here bear a Dirac matrix,
\begin{equation}
\left.\Lagr^{\mathrm{ferm}}_{\mathrm{eff}}\right|_{\mathcal{O}(m^2)}=-i\int\frac{\dd^d q}{(2\pi)^d}\int^m\dd m'\,\tr\,\left(\Delta\gamma^\mu\Delta\gamma^\nu\Delta\dc_\mu\dc_\nu + \Delta\gamma^\mu\Delta\gamma^\nu (\dc_\mu\dc_\nu\Delta)\right)\, .
\label{Lfermm2}
\end{equation}
We split the propagators according to Eq.~\eqref{DeltaFermSplit}, we denote $\Delta_f=-\slashed q/(q^2-m'^2)$ and $\Delta_b=m'/(q^2-m'^2)$, and then commute the momentum dependence to the left. Since the integration with an odd power in $q$ in the numerator vanishes, we can only have an even power in $\Delta_f$. We then perform the integration in terms of the fermionic master integrals.

An extra step that arises in the fermionic CDE is to perform the Dirac trace, or at least simplify the contractions in-between Dirac matrices, in order to be able to form covariant operators. Extra care must be taken for the terms that bear open covariant derivatives to the right since they carry the spin-connection.

Let's focus on the first term of Eq.~\eqref{Lfermm2}. After splitting the propagators we obtain,
\begin{align}
\begin{split}
&-i\int\frac{\dd^d q}{(2\pi)^d}\int^m\dd m'\,\tr\,\Big((\Delta_b\gamma^\mu\Delta_b\gamma^\nu\Delta_b+\Delta_f\gamma^\mu\Delta_f\gamma^\nu\Delta_b\\
&\quad\quad\quad\quad+\Delta_f\gamma^\mu\Delta_b\gamma^\nu\Delta_f+\Delta_b\gamma^\mu\Delta_f\gamma^\nu\Delta_f)\dc_\mu\dc_\nu \Big)\\
&=i\g\,\tr\,\Bigg(\bigg(\mathcal{K}[q^0]^3_3\gamma^\mu\gamma^\nu+\mathcal{K}[q^2]^1_3g_{\alpha\beta}\left( \gamma^\alpha\gamma^\mu\gamma^\beta\gamma^\nu +\gamma^\alpha\gamma^\mu\gamma^\nu\gamma^\beta+\gamma^\mu\gamma^\alpha\gamma^\nu\gamma^\beta\right)\bigg)\dc_\mu\dc_\nu\Bigg)\, .
\label{FermExample1}
\end{split}
\end{align}
The first possibility is to simplify the contractions among Dirac matrices using the Clifford algebra, and then form covariant quantities,
\begin{align}
\begin{split}
&i\g\,\tr\,\Bigg( \mathcal{K}[q^0]^3_3\sdc^2 + d\,\mathcal{K}[q^2]^1_3\sdc^2-4\mathcal{K}[q^2]^1_3\dc^2 \Bigg)\\
&=\g\frac{-1}{16\pi^2}\frac{m^2}{2}\left(1-\lm\right)\tr\,\left( \sdc^2-\dc^2 \right)\\
&=\g\frac{-1}{16\pi^2}\frac{m^2}{2}\left(1-\lm\right)\tr\,\left( -\frac{R}{4} \right)\, .
\end{split}
\end{align}
The remaining trace is over gauge and spin indices. From the first to the second line we discarded the pole $2/\bar\epsilon$, although we took care not to forget the finite contribution obtained when multiplied by $d=4-\epsilon$ in the first line. From the second to the last line we used Eq.~\eqref{eq:sigmaFgrav}.

Another possibility is, from Eq.~\eqref{FermExample1}, to make explicit the spin-connection, directly compute the Dirac trace, and then form covariant quantities with the explicit spin-connections. When the open derivatives are on the far right we can write: $\dc_\mu\dc_\nu=(\pa_\mu \omega_\nu)+\omega_\mu\omega_\nu$. In RNC it takes a simple form as explained in Appendix~\ref{Appendix:RNCFS},
\begin{equation}
\dc_\mu\dc_\nu=-\frac{1}{8}\gamma^\alpha\gamma^\beta R_{\nu\mu\alpha\beta}\, .
\end{equation}
Using the RNC for the spin-connection is the fastest method to get covariant quantities when the computation involves more terms.

The systematic procedure is the same as for the bosonic CDE: commute momentum dependence to the left, perform the mass and momentum integration, then form covariant quantities. The last step is slighty more involved for the fermionic CDE, but with the use of RNC it is straightforward.

\section{Bosonic UOLEA in curved spacetime}
\label{sec4}

As opposed to the fermionic CDE in curved spacetime, the results from the bosonic CDE presented here are well-known. Indeed our results can be matched for example with those from the heat kernel approach \cite{Avramidi:2000bm} \footnote{Note that due to the difference in conventions in the one-loop effective action of \cite{Avramidi:2000bm}, we have $m^2_{\mathrm{HK}}=-m^2_{\mathrm{CDE}}$ and $S^{\mathrm{CDE}}_{\mathrm{eff}}=2S^{\mathrm{HK}}_{\mathrm{eff}}$.}, or more recently using the worldline formalism \cite{Bastianelli:2008cu}. Nevertheless, the CDE has the advantage of being systematic and in fact algorithmic, thus the expansion is easy to automatise. In addition, the CDE being based on an inverse mass expansion its physical interpretation is always enlightening compared to a quite formal heat kernel approach.

The CDE in curved spacetime has already been approached in \cite{Binetruy:1988nx,Alonso:2019mok}. However, as opposed to our current method, these works use the Gaillard and Cheyette sandwish mentionned in Section \ref{BosUOLEAflat}. As explained earlier, it has the advantage of making the computation manifestly covariant, but at the cost of making the expansion significantly more complex. When the curvature of spacetime is introduced, such computation can quickly become untractable. By avoiding this step, we make the computation simpler which allows us to compute higher order corrections in a straightforward way. The computation of non-renormalisable operators (order $1/m^2$) using the CDE in curved spacetime are presented for the first time. More than that, our systematic method could easily be implemented in a code which would allow to generate even higher dimensional operators associated to generic UV theories involving gravity. The drawback of our method would then be to form the covariant operators at the end, but thanks to the RNC and the FS gauge (see Appendix~\ref{Appendix:RNCFS} for additional details) it turns out to be straightforward and algorithmic as well. Another advantage of our method is that it is independent of a choice of coordinate system, as opposed to \cite{Binetruy:1988nx}, therefore it can be used to compute non-covariant quantities such as consistent gravitational anomalies (see for example \cite{Filoche:2022dxl}).

Note that the result from this section can apply to the integration of a (real or complex) scalar, a vector-like fermion, massive and massless vector bosons, as well as ghosts \cite{Henning:2014wua}. The factor $c_s$ and the content of $U$ depend on the nature of the heavy field. One significant asset of our approach is that one can straightforwardly incorporate all the improvements on the CDE in flat spacetime EFTs, such as non-degenerate mass matrix\cite{Drozd:2015rsp}, mixed heavy light loops\cite{Ellis:2016enq} and UV theories involving derivative couplings \cite{Quevillon:2021sfz}~\footnote{The covariant diagrams~\cite{Zhang:2016pja} can also be used to enumerate the terms of the expansion, but it does not account for the commutation of the momentum dependence to the left of the derivatives so most of their properties must be dropped.}.\\

We will now compute the terms of order $m^2$ and $m^0$, which consist in renormalisable operators.
We recall the bosonic one-loop effective action,
\begin{align}
\begin{split}
S^{\mathrm{bos}}_{\mathrm{eff}}&=ic_s\Tr\log(\dc^2+m^2+U)\\
&=-ic_s\int \dd^dx \frac{\dd^d q}{(2\pi)^d}\int^{m^2} \dd m'^2\tr\,\sum_{n\geq 0}\left[\Delta(g^{\mu\nu}D_\mu D_\nu-g^{\mu\nu}\Gamma^\rho_{\mu\nu}D_\rho+2iq\cdot D-ig^{\mu\nu}\Gamma^\rho_{\mu\nu}q_\rho+U)\right]^n\Delta\, .
\label{Expansion2}
\end{split}
\end{align}
The terms that contribute at order $m^2$ are,
\begin{equation}
\left.\Lagr^{\mathrm{bos}}_{\mathrm{eff}}\right|_{\mathcal{O}(m^2)}=-ic_s\int \frac{\dd^d q}{(2\pi)^d}\int \dd m'^2\,\tr\,\left[\Delta (D^2-g^{\mu\nu}\Gamma^\rho_{\mu\nu}D_\rho+U)\Delta+\left(\Delta (2iq\cdot D-ig^{\mu\nu}\Gamma^\rho_{\mu\nu}q_\rho)\right)^2\right]\Delta\, .
\label{m2Exp}
\end{equation}

As explained in Section \ref{SystematicProcedure}, we first commute the momentum dependence to the left, and the covariant derivatives to the right. We then perform the integration in terms of master integrals, and in RNC, we directly obtain,
\begin{equation}
\left.\Lagr^{\mathrm{bos}}_{\mathrm{eff}}\right|_{\mathcal{O}(m^2)}=\frac{c_s}{16\pi^2}\g m^2\left(1-\lm\right)\tr\left(\,\frac{1}{6}R+U\right)\, .
\label{m2scalar}
\end{equation}

We are now interested in the terms of order $m^0$. For simplicity we note $\Gamma D=g^{\mu\nu}\Gamma^\rho_{\mu\nu}D_\rho$ and $\Gamma q=ig^{\mu\nu}\Gamma^\rho_{\mu\nu}q_\rho$.
In RNC, the Christoffel symbols locally vanish unless a derivative acts on them. Therefore the terms that contribute must have enough covariant derivative to the left of the Christoffel symbols.
The operators that contribute in RNC are listed in Tab.~\ref{tab2}.

\begin{table}[h!]
\begin{center}
\begin{tabular}{ |c|m{8cm}| } 
  \hline
  $n=2$ & $$\tr\,\Delta (D^2 +U)\Delta (D^2-\Gamma D+U)\Delta$$ \\ 
  \hline
  $n=3$ & $$\tr\,\Delta (D^2 +U)\left(\Delta (2iq\cdot D-\Gamma q)\right)^2\Delta\, ,$$ $$\tr\,\Delta 2iq\cdot D \Delta (D^2-\Gamma D +U)\Delta(2iq.\cdot D -\Gamma q)\Delta\, ,$$ $$\tr\,\Delta 2iq\cdot D\Delta(2iq\cdot D-\Gamma q)\Delta (D^2-\Gamma D +U)\Delta\, .$$ \\ 
  \hline
  $n=4$ & $$\tr\,[\Delta 2iq\cdot D]^4\Delta\, ,$$ $$\tr\, \left[\Delta 2iq\cdot D \Delta (-\Gamma q)\right]^2\Delta\, ,$$ $$\tr\, [\Delta 2iq\cdot D]^2\left[\Delta (-\Gamma q)\right]^2\Delta\, ,$$ $$\tr\,[\Delta 2iq\cdot D]^3\Delta(-\Gamma q)\Delta\, ,$$ $$\tr\, [\Delta 2iq\cdot D]^2\Delta(-\Gamma q)\Delta 2iq\cdot D\Delta\, ,$$ $$\tr\, 2iq\cdot D\Delta(-\Gamma q)[\Delta 2iq\cdot D]^2\Delta\, . $$\\
  \hline
\end{tabular}
\end{center}
\caption{Operators that contribute at order $m^0$ in RNC.}
\label{tab2}
\end{table}

For each of these terms, we follow the usual procedure: we commute the momentum dependence to the left using Table~\ref{tab1}, it is helpful to form covariant quantities later to fully commute the covariant derivatives to the right. We then perform the integration over momentum and mass using the master integrals. Finally we can use the RNC formulae provided in App.~\ref{Appendix:RNCFS} to make the covariant curvature quantities appear. As for the covariant derivatives, they have to be combined together to form covariant quantities as well, at order $m^0$ is it straightforward so the use of FS gauge is not necessary.

The RNC expansions are required up to order four in the metric, and three in the Christoffels. For example we have in RNC,
\begin{align}
(D_\mu D_\nu D_\rho D_\sigma \Delta)=&-(\pa_{\mu\nu\rho\sigma}g^{\alpha\beta})q_\alpha q_\beta \Delta^2\nonumber\\
+&2\left(
(\pa_{\mu\nu}g^{\alpha\beta})(\pa_{\rho\sigma}g^{\gamma\delta}) + (\pa_{\mu\rho}g^{\alpha\beta})(\pa_{\nu\sigma}g^{\gamma\delta}) + (\pa_{\mu\sigma}g^{\alpha\beta})(\pa_{\nu\rho}g^{\gamma\delta})
\right)q_\alpha q_\beta q_\gamma q_\delta \Delta^3\, .
\end{align}
Finally, after using Bianchi identities, we simply obtain the following terms of order $m^0$,
\begin{align}
\begin{split}
\left.\Lagr^{\mathrm{bos}}_{\mathrm{eff}}\right|_{\mathcal{O}(m^0)}=\frac{c_s}{16\pi^2}\g\lm \tr\,\bigg[&-\frac{1}{72}R^2+\frac{1}{180}R_{\mu\nu}R^{\mu\nu}-\frac{1}{180}R_{\mu\nu\rho\sigma}R^{\mu\nu\rho\sigma}\\
&-\frac{1}{30}(\Box R)\\
&-\frac{1}{6}R U-\frac{1}{6}(\Box U)-\frac{1}{2}U^2-\frac{1}{12}F^2 \bigg]\, ,
\label{bosm0}
\end{split}
\end{align}
where $\Box=\dc^2$.

Note that since we performed the expansion with the covariant derivatives in the field representation $D$, we obtain at first $g^{\mu\nu}(D_\mu D_\nu U)$ which is not diffeomorphism covariant. However, in RNC it is equal to $(\Box U)$ which is diffeomorphism covariant. If we had performed the expansion without RNC, we would obtain some non-covariant operators to combine with $g^{\mu\nu}(D_\mu D_\nu U)$ to form $(\Box U)$.

Note also that if we had performed the expansion keeping the general covariant derivative $\dc$, we would obtain $\tr\,\mathcal{F}^2$ instead of $\tr\,F^2$. However, they are equal since $\tr\,\mathcal{F}^2=\tr\,\mathcal{F}_{\mu\nu}F^{\mu\nu}=\tr\,\left(F^2+R_{\mu\nu}F^{\mu\nu}-R_{\mu\nu}F^{\mu\nu}\right)=\tr\,F^2$.

We should comment on the fact that the terms of order $m^2$ and $m^0$ are divergent and these divergences can be absorbed by the renormalisation as it is well known. We used dimensional regularisation to compute the divergent momentum integrals with $\overline{MS}$ scheme, and $\mu$ is the renormalisation scale. In practice, these contributions can conveniently be used to compute the RGE running of the EFT operators at one-loop (see \cite{Henning:2016lyp} for example).\\

We now outline the computation of the terms of order $1/m^2$, which consists in the first order of non-renormalisable operators. For simplicity, we assume a weak gravitational field, so we only consider term that are linear in curvature~\footnote{There is no conceptual difficulty in obtaining the terms more than linear in gravity, the only difference is that the next order of the expansion in RNC of the metric and the Christoffel symbols are required. We are also interested in recovering the seminal result from \cite{Drummond:1979pp}.}. We collect in Table~\ref{tab3} the terms of order $1/m^2$, discarding the terms that are not linear in curvature, and that vanish in RNC. The terms that are linear in curvatures arise from second derivatives of the propagators $\Delta$ (hence second derivative of the metric), and first derivative of the Christoffel connection.

\begin{table}[h!]
\begin{center}
\begin{tabular}{ |c| m{12cm}| } 
  \hline
  $n=3$ & $$\tr\,\Delta (D^2 +U)\Delta (D^2-\Gamma D+U)\Delta(D^2-\Gamma D +U)\Delta$$ \\ 
  \hline
  $n=4$ & $$\tr\,[\Delta (D^2-\Gamma D +U)]^2 [\Delta (2iq\cdot D-\Gamma q)]^2 \Delta$$ $$\tr\,\Delta (D^2 +U) [\Delta (2iq\cdot D-\Gamma q)]^2  \Delta (D^2-\Gamma D +U) \Delta$$ $$\tr\,[\Delta (2iq\cdot D-\Gamma q)]^2 [\Delta (D^2-\Gamma D +U)]^2\Delta$$ $$\tr\,\Delta (D^2 +U) \Delta (2iq\cdot D-\Gamma q) \Delta (D^2-\Gamma D +U) \Delta (2iq\cdot D-\Gamma q) \Delta$$  $$\tr\,\Delta (2iq\cdot D) \Delta (D^2-\Gamma D +U) \Delta (2iq\cdot D-\Gamma q) \Delta (D^2-\Gamma D +U) \Delta$$ $$\Delta (2iq\cdot D) [\Delta (D^2-\Gamma D +U)]^2 \Delta (2iq\cdot D-\Gamma q) \Delta$$  \\ 
  \hline
  $n=5$ & $$\tr\,\Delta (D^2 +U) \Delta [\Delta (2iq\cdot D-\Gamma q)]^4 \Delta$$ $$\tr\,\Delta (2iq\cdot D) \Delta (D^2-\Gamma D +U) [\Delta (2iq\cdot D-\Gamma q)]^3 \Delta$$ $$\tr\,[\Delta (2iq\cdot D-\Gamma q)]^2 \Delta (D^2-\Gamma D +U) [\Delta (2iq\cdot D-\Gamma q)]^2 \Delta$$ $$\tr\,[\Delta (2iq\cdot D-\Gamma q)]^3 \Delta (D^2-\Gamma D +U) [\Delta (2iq\cdot D-\Gamma q)] \Delta$$ $$\tr\,[\Delta (2iq\cdot D-\Gamma q)]^4 \Delta (D^2-\Gamma D +U) \Delta$$ \\ 
  \hline
  $n=6$ & $$\tr\,[\Delta (2iq\cdot D-\Gamma q)]^6 \Delta$$\\
  \hline
\end{tabular}
\end{center}
\caption{Operators that contribute at order $1/m^2$ in RNC, and linear in curvature.}
\label{tab3}
\end{table}

Commuting in RNC the momentum dependence to the left, and performing the integration over mass then momentum, we obtain terms which are separately gauge variant. The next step is to form covariant quantities.

Finally we have terms of the form $D^6$, they have to be combined together again in order to make the gauge invariant form explicit. There are two possibilities to easily form covariant operators, the first one is to identify the covariant basis and then solve a system to change basis, the second one is to use the Fock-Schwinger gauge (see App.\ref{Appendix:RNCFS}).
    
After recasting the operators in a covariant form, we simply obtain,
\begin{align}
\begin{split}
\left.\Lagr^{\mathrm{bos}}_{\mathrm{eff}}\right|_{\mathcal{O}(1/m^2)}=\frac{c_s}{16\pi^2}\g\frac{1}{m^2} \tr\,\bigg[& -\frac{1}{72}RF^2-\frac{1}{90}R_{\mu\nu}F^{\mu\lambda}\tensor{F}{^\nu_\lambda}-\frac{1}{180}R_{\mu\nu\rho\sigma}F^{\mu\nu}F^{\rho\sigma}\\
& +\left(\frac{1}{90}-\frac{a}{2}\right)(\dc_\mu F^{\mu\nu})^2+a\tensor{F}{_\mu^\nu}\tensor{F}{_\nu^\rho}\tensor{F}{_\rho^\mu}+\left( \frac{1}{360}+\frac{a}{4} \right)\left(\dc_\mu F_{\nu\rho}\right)\left(\dc^\mu F^{\nu\rho}\right) \\
& -\frac{1}{12}U(\Box U) -\frac{1}{36}R(\Box U) - \frac{1}{12}RU^2-\frac{1}{6}U^3-\frac{1}{12}UF^2 \bigg]\, ,
\label{1/m2term}
\end{split}
\end{align}
where $\Box=\dc^2$. The result is independent of the parameter $a$. It bears the freedom of the choice of covariant basis since $\left(\dc_\mu F_{\nu\rho}\right)\left(\dc^\mu F^{\nu\rho}\right)$, $\tensor{F}{_\mu^\nu}\tensor{F}{_\nu^\rho}\tensor{F}{_\rho^\mu}$ and $(\dc_\mu F^{\mu\nu})^2$ are related by integrations by parts and Bianchi identity.

As far as we know, this is the first time that this contribution to the one-loop effective action, corresponding to non-renormalisable operators, have been derived via the CDE methods in curved spacetime. 

The result we obtain is invariant since the operator basis,
\begin{equation}
    B_{\dc,F}=\{\left(\dc_\mu F_{\nu\rho}\right)\left(\dc^\mu F^{\nu\rho}\right),\tensor{F}{_\mu^\nu}\tensor{F}{_\nu^\rho}\tensor{F}{_\rho^\mu},(\dc_\mu F^{\mu\nu})^2\}\, ,
\end{equation}
is both diffeomorphism and gauge invariant. We recall that $F$ is the gauge and spin-connection field strength, while $\dc$ bears the gauge connection and spin-connection as well as the Christoffel connection.
Note that the basis $B_{D,F}=\{\left(D_\mu F_{\nu\rho}\right)\left(D^\mu F^{\nu\rho}\right),\tensor{F}{_\mu^\nu}\tensor{F}{_\nu^\rho}\tensor{F}{_\rho^\mu},(D_\mu F^{\mu\nu})^2\}$ is not diffeomorphism invariant, and the basis $B_{\dc,\mathcal{F}}=\{\left(\dc_\mu \mathcal{F}_{\nu\rho}\right)\left(\dc^\mu \mathcal{F}^{\nu\rho}\right),\tensor{\mathcal{F}}{_\mu^\nu}\tensor{\mathcal{F}}{_\nu^\rho}\tensor{\mathcal{F}}{_\rho^\mu},(\dc_\mu \mathcal{F}^{\mu\nu})^2\}$ is not gauge invariant. 

The expansion was performed with the covariant derivatives $D$ (i.e Eq.~\eqref{MasterFormula2}). Therefore, when forming the invariant operators, we actually obtain at first the basis $B_{D,F}$. However in RNC it is equal to $B_{\dc,F}$. If the computation were performed without RNC, we would obtain terms which are not diffeomorphism invariant, to combine with the operators from the basis $B_{D,F}$ to form operators from the basis $B_{\dc,F}$. Likewise, we obtain at first $g^{\mu\nu}(D_\mu D_\nu U)$, which in RNC is equal to $g^{\mu\nu}(\dc_\mu\dc_\nu U)=(\Box U)$.

Similarly, if we employed the expansion keeping the general covariant derivatives $\dc$ (i.e Eq.~\eqref{MasterFormula1}), we would obtain the basis $B_{\dc,\mathcal{F}}$ which has to be combined with the terms that are not gauge invariant to form operators from the basis $B_{\dc,F}$.

However, if we use the FS gauge to form the covariant quantities, we directly obtain the correct basis $B_{\dc,F}$ (see App.~\ref{Appendix:RNCFS}).

The derivatives that are localised on curvatures are independent of this choice of using $\dc$ (i.e Eq.~\eqref{MasterFormula1}) or $D$ (i.e Eq.~\eqref{MasterFormula2}) since they arise from the RNC formulae.\\

Finally, the bosonic universal one-loop effective action in curved spacetime, up to dimension 6 operators, reads,
\begin{align}
\begin{split}
S^{\mathrm{bos}}_{\mathrm{eff}}=\frac{c_s}{16\pi^2}&\int\g\dd^4 x\, \tr\,\Bigg\{m^2\left(1-\lm\right)\left(\,\frac{1}{6}R+U\right)\\
&\\
&+\lm \bigg[-\frac{1}{72}R^2+\frac{1}{180}R_{\mu\nu}R^{\mu\nu}-\frac{1}{180}R_{\mu\nu\rho\sigma}R^{\mu\nu\rho\sigma}-\frac{1}{30}(\Box R)\\
&\hspace{0,15\textwidth}-\frac{1}{6}R U-\frac{1}{6}(\Box U)-\frac{1}{2}U^2-\frac{1}{12}F^2\bigg]\\
&\\
&+\frac{1}{m^2}\bigg[ -\frac{1}{72}RF^2-\frac{1}{90}R_{\mu\nu}F^{\mu\lambda}\tensor{F}{^\nu_\lambda}-\frac{1}{180}R_{\mu\nu\rho\sigma}F^{\mu\nu}F^{\rho\sigma}\\
& \hspace{0.08\textwidth}+\left(\frac{1}{90}-\frac{a}{2}\right)(\dc_\mu F^{\mu\nu})^2+a\tensor{F}{_\mu^\nu}\tensor{F}{_\nu^\rho}\tensor{F}{_\rho^\mu}+\left( \frac{1}{360}+\frac{a}{4} \right)\left(\dc_\mu F_{\nu\rho}\right)\left(\dc^\mu F^{\nu\rho}\right) \\
& \hspace{0.08\textwidth}-\frac{1}{12}U(\Box U) -\frac{1}{36}R(\Box U) - \frac{1}{12}RU^2-\frac{1}{6}U^3-\frac{1}{12}UF^2 +\mathcal{O}(R^2)\bigg]\\
&+\mathcal{O}(1/m^4)
\Bigg\}\, .
\label{UOLEAgravBos}
\end{split}
\end{align}
The result is independent of $a$, and only the terms linear in curvature were computed at order $1/m^2$. The remaining trace is over gauge and spin indices.
This result is in agreement with \cite{Avramidi:2000bm}.

\subsection{Integrating out the graviton}

So far the computation were performed on a fixed spacetime background. It is of interest to treat the metric as a dynamical field, and integrate over its configurations. As we would do for a spin-1 gauge field, we use the background field method: the metric is split as $g+\delta g$, where $g$ is a fixed background, and $\delta g$ is a fluctuation integrated over in the path integral. Indices are lowered and raised with the background metric. The quantum field $\delta g_{\mu\nu}$ inherites a gauge invariance from the diffeomorphism invariance of the background, which is dealt with using the Fadeev-Popov procedure. The action is then expanded up to quadratic order in $\delta g$ so that the path integral can be performed \footnote{The use of the saddle point approximation is only possible around a background $g$ that is a saddle point of the space of metric configurations. We assume $g$ fulfills that requirement. The complete gravitational path integral would then be the sum of the saddle point contributions, supplemented by other non-perturbative configurations, such as instantons, which cannot be treated following the presented procedure. For recent literature see for example \cite{Castro:2011xb,Castro:2023dxp,Benjamin:2023uib}}.

There is however a discrepancy when dealing with a spin-2 field with respect to fields of smaller spin: the field-space has a non-trivial metric $G$. A consequence is that the second derivative of the action is not a scalar under a redefinition of $\delta g$. This issue was solved in \cite{Vilkovisky:1984st} by introducing a field-space covariant derivative. For a generic field $\phi$ with kinetic term $\dot\phi_a A^{ab}(\phi)\dot\phi_b$, the field-space metric is $G^{ab}(x,y)=A^{ab}(\phi)\delta(x-y)$ and  the connection is $\hat\Gamma^a_{bc}=\frac{1}{2}(G^{-1})^{ad}(\frac{\delta G_{bd}}{\delta \phi^c}+\frac{\delta G_{cd}}{\delta \phi^b}-\frac{\delta G_{bc}}{\delta \phi^d})$. We can see right away that if the action is at most quadratic in $\phi$, then $A^{ab}$ is independent of $\phi$ and $\hat\Gamma$ vanishes. Therefore, for spin 0, $1/2$, 1 and $3/2$ matter fields, the second derivative of the action is a scalar and no field-space covariant derivative is needed.

The one-loop effective action is covariantly defined as,
\begin{equation}
iS_{\mathrm{eff}[g]}=\log\int\sqrt{G}\left[\dc\delta g\right]e^{iS[g]+\frac{i}{2}\delta g(\hat\dc^2 S)[g]\delta g+\mathcal{O}(\delta g^3)}\simeq iS[g]-\frac{1}{2}\Tr\log\left(-(\hat\dc^2 S)[g]\cdot G^{-1}\right)\, ,
\label{covEFT}
\end{equation}
where the background $g$ is solution to the equations of motion, that is $(\hat\dc S)[g]=\frac{\delta S}{\delta g_{\mu\nu}}[g]=0$.
$\hat\dc$ is the field-space covariant derivative such that \footnote{We leave implicit the summation over spacetime indices in the second term. $(\hat D^2 S)(x,y)$ remains proportional to $\delta^d(x-y)$.},
\begin{equation}
(\hat\dc^2 S)=\frac{\delta^2 S}{\delta g_{\mu\nu}\delta g_{\rho\sigma}}+\hat\Gamma^{\mu\nu,\rho\sigma}_{\alpha\beta}\frac{\delta S}{\delta g_{\alpha\beta}}\, .
\label{D^2 S}
\end{equation}
The field-space metric and its inverse are,
\begin{equation}
G^{\mu\nu,\rho\sigma}=\frac{1}{4}\left( g^{\mu \rho}g^{\nu\sigma}+g^{\mu \sigma}g^{\nu\rho}-g^{\mu\nu}g^{\rho\sigma} \right)\quad\text{and, }\quad (G^{-1}_{\mu\nu,\rho\sigma})=g_{\mu\rho}g_{\nu\sigma}+g_{\mu \sigma}g_{\nu\rho}-g_{\mu\nu}g_{\rho\sigma}\, .
\end{equation}
Eq.~\eqref{D^2 S} evaluated on $g$ solution to the equations of motion reduces to the usual second derivative of the action. But if one wanted to vary the background $g$, then the variation of $\hat \dc^2 S$ would cancel exactly against the variation of $\sqrt{G}$ from the measure, making the theory invariant under a redefinition of $\delta g$.

We consider the UV theory,
\begin{equation}
S=\int\dd^4 x\g\left(\frac{1}{4\kappa}\left(2\Lambda-R\right)+\Lagr_{\mathrm{mat}}\right)\, ,
\end{equation}
where $\kappa=8\pi/M_P^2$ with $M_P$ the Planck mass, and $\Lagr_{\mathrm{mat}}$ is the matter Lagrangian.
With the background $g$ on-shell, the second derivative of the action with respect to the metric, including the ghosts $c^\mu$ and $\bar c^\mu$, and the gauge-fixing in harmonic gauge \footnote{Note that it corresponds to the gauge invariance associated to $\delta g$. We still have the freedom to choose the coordinate system, such as the RNC.}, reads,
\begin{align}
\begin{split}
\int\dd^4 x \frac{1}{2}\delta g\frac{\delta^2 S}{\delta g^2}\delta g=&-\int\dd^4x\g\,\bar c^\mu(g_{\mu\nu}\dc^2+R_{\mu\nu})c^\nu\\
&-\int\dd^4x\g\,\delta g_{\alpha\beta}\left(\frac{1}{2}g^\alpha_{(\rho}g^\beta_{\sigma)}\dc^2+\tensor{R}{^\alpha_{(\rho}^\beta_{\sigma)}}-g^{\alpha\beta}R_{\rho\sigma}+\Lambda g^{\alpha\beta}g_{\rho\sigma}\right)G^{\gamma\delta,\rho\sigma}\delta g_{\gamma\delta}\\
&-\int\dd^4 x\g \delta g_{\alpha\beta }\tensor{\mathcal{O}}{^\alpha^\beta_\rho_\sigma}G^{\gamma\delta,\rho\sigma}\delta g_{\gamma\delta}\, ,
\label{d2gS}
\end{split}
\end{align}
where $\mathcal{O}$ corresponds to the matter part. The parenthesis around the indices denotes the symmetrisation: $T_{(\mu\nu)}=T_{\mu\nu}+T_{\nu\mu}$. Including loops of graviton only, $\mathcal{O}$ is a local operator and reads,
\begin{equation}
\mathcal{O}\cdot G=\frac{\kappa}{\g}(\hat\dc\g T)\, ,
\end{equation}
with $T$ the matter energy-momentum tensor such that $\g T=-2\frac{\delta S}{\delta g}$. More details about the derivation can be found in \cite{Donoghue:1995cz,Alonso:2019mok}.
In practice, $\mathcal{O}$ can be computed by taking the second derivative of the matter action with respect to the metric.

According to Eq.~\eqref{covEFT}, the graviton piece must be contracted with $G^{-1}$, which yields the identity when contracted against $G$ from Eq.~\eqref{d2gS}. Finally, we obtain the one-loop effective action after integrating out the graviton (first line) and ghost (second line),
\begin{align}
S^{\mathrm{graviton}}_{\mathrm{eff}}=&\frac{i}{2}\Tr\log\left(\frac{1}{2}g^\mu_{(\rho}g^\nu_{\sigma)}\dc^2+\tensor{R}{^\mu_{(\rho}^\nu_{\sigma)}}-g^{\mu\nu}R_{\rho\sigma}+\Lambda g^{\mu\nu}g_{\rho\sigma}+\tensor{\mathcal{O}}{^\mu^\nu_\rho_\sigma}\right)\nonumber\\
&-i\Tr\log\left( g_{\mu\nu}\dc^2+R_{\mu\nu} \right)\, ,
\label{gravitonOneLoop}
\end{align}
Both are of the form $\Tr\log\left( \mathds{1}\dc^2+U \right)$, therefore they can be obtained from the bosonic UOLEA derived previously.
Note that the trace is also performed over Lorentz indices. The graviton piece is traced over by contracting the indices with the identity for order 4 Lorentz tensors,
\begin{equation}
\mathds{1}^{\mu\nu}_{\rho\sigma}=G^{\mu\nu,\alpha\beta}(G^{-1})_{\alpha\beta,\rho\sigma}=\frac{1}{2}g^\mu_{(\rho}g^\nu_{\sigma)}\, .
\end{equation}

Note that since the ghost and the graviton are massless, the effective action contains infrared divergences. They can be regulated by inserting a mass term $\dc^2\to\dc^2+m^2$ \cite{Henning:2014wua}.
For loops of graviton only, $\mathcal{O}$ is a local operator and we can apply the results from the bosonic UOLEA Eq.~\eqref{UOLEAgravBos}, with,
\begin{align}
&c_s^{\mathrm{graviton}}=1/2\, \text{, }\,U_{\mathrm{graviton}}=\tensor{R}{^\mu_{(\rho}^\nu_{\sigma)}}-g^{\mu\nu}R_{\rho\sigma}+\Lambda g^{\mu\nu}g_{\rho\sigma}+\tensor{\mathcal{O}}{^\mu^\nu_\rho_\sigma}\\
&c_s^{\mathrm{ghost}}=-1\, \text{, }\,\hspace{0,45cm}U_{\mathrm{ghost}}=R_{\mu\nu}\, .
\end{align}

Note that if mixed species loops are allowed, then the bosonic UOLEA does not apply to the graviton since $\mathcal{O}$ includes open derivatives, the CDE has to be performed from scratch following the mixed heavy-light methods \cite{Ellis:2016enq,Ellis:2017jns,Alonso:2019mok}.

\section{Fermionic UOLEA in curved spacetime}
\label{sec5}
We now turn to the fermionic CDE in curved spacetime.
The one-loop effective action that we obtain after integrating out a fermion is of the form,
\begin{equation}
S^{\mathrm{ferm}}_{\mathrm{eff}}=-i\Tr\log\,\left(\g\left(i\sdc-m-Q\right)\right)\, ,
\label{FermDet}
\end{equation}
In a similar manner as the bosonic determinant, it can be expanded as,
\begin{align}
S^{\mathrm{ferm}}_{\mathrm{eff}}=i\int\dd^dx\frac{\dd^dq}{(2\pi)^d}\int^m \dd m'\,\tr\,\sum_{n\geq0}\left[\Delta\left( -i\sdc+Q \right)\right]^n\Delta\, ,
\end{align}
where $\Delta=m'/(q^2-m'^2)-\slashed q/(q^2-m'^2)$.
The mass order is rather simple, since the $n$-th term of the sum is proportional to $m^{4-n}$.

We can split $Q=Q_e+Q_o$ where $Q_e$ (resp. $Q_o$) has an even (resp. odd) number of Dirac matrices. Following \cite{Ellis:2020ivx}, we can assume the general form,
\begin{equation}
Q_e=W_o+i W_1\gamma_5,\quad Q_o=X_\mu\gamma^\mu+i A_\mu\gamma^\mu\gamma_5\, .
\end{equation}
We choose to limit ourselves to the case of a scalar, pseudo-scalar, vector and pseudo-vector term, but the computation can be performed for any type of operator. 

For a chiral fermion, it is simpler to keep $(\dc_\mu\psi)=(\pa_\mu+iV_\mu+\omega_\mu)\psi$ where $\omega$ is the spin-connection and $V$ a vector gauge field, and put the axial field in $Q_o\supset -\slashed A\gamma_5$.

To compute the traces, we use the Breitenlohner-Maison-’t Hooft-Veltman (BMHV) scheme for $\gamma_5$ \cite{tHooft:1972tcz,Breitenlohner:1977hr}. The choice of scheme for $\gamma_5$ can have consequences which should not be disregarded, especially in the context of quantum anomalies \cite{Filoche:2022dxl}.

We emphasise again that others approaches to compute one-loop effective actions in gravity, such as heat kernel, CDE and worldline formalism \cite{Avramidi:2000bm,Binetruy:1988nx,Alonso:2019mok,Bastianelli:2008cu,Davila:2009vt}, always applied to a bosonic determinant, therefore were restrained to $Q=W_0+\gamma^\mu X_\mu$, as explained in Section \ref{UOLEAferm}. To our knowledge, the fermionic Universal One-Loop Effective Action on a general spacetime manifold were never computed before.

Improvements on the CDE can also be applied (namely, non-degenerate mass matrix\cite{Drozd:2015rsp}, mixed heavy-light \cite{Ellis:2016enq}). The covariant diagrams can be used to enumerate the terms of the expansion \cite{Zhang:2016pja}, but it does not account for the commutation of the momentum dependence to the left of the derivatives so most of their properties must be dropped.

\subsection{Effective action operators}

At order $m^3$, discarding the terms with odd number of Dirac matrices which vanish under the trace, we have,
\begin{equation}
\left.S^{\mathrm{ferm}}_{\mathrm{eff}}\right|_{\mathcal{O}(m^3)}=i\int\dd^dx\frac{\dd^dq}{(2\pi)^d}\int^m \dd m'\,\tr \Delta Q_e\Delta \, .
\end{equation}
There are no open derivatives, so the momenta can be commuted trivially to the left, and after integration we obtain,
\begin{equation}
\left.\Lagr^{\mathrm{ferm}}_{\mathrm{eff}}\right|_{\mathcal{O}(m^3)}=i\g\,\tr\,\left( \mathcal{K}[q^0]^2_2 Q_e + \mathcal{K}[q^2]^0_2 g_{\mu\nu}\gamma^\mu Q_e \gamma^\nu \right)
\label{m3ferm1}
\end{equation}

So far we have not performed the traces, nor have we use trace cyclicity or anticommutation relation between $\gamma_5$ and a Dirac matrix. The result in Eq.~\eqref{m3ferm1} is therefore independent of the choice of scheme for $\gamma_5$ in dimensional regularisation. 
From Eq.~\eqref{m3ferm1} we need to explicit $Q_e=W_o+iW_1\gamma_5$ to perform the trace using the BMHV scheme for $\gamma_5$. Discarding the poles we obtain,
\begin{equation}
\left.\Lagr^{\mathrm{ferm}}_{\mathrm{eff}}\right|_{\mathcal{O}(m^3)}=\g\frac{-1}{16\pi^2}4m^3\left(1-\lm\right)\tr\, W_0\, .
\end{equation}
The remaining trace is over gauge degrees of freedom.\\

The non-vanishing contributions at order $m^2$ are,
\begin{align}
\begin{split}
\left.S^{\mathrm{ferm}}_{\mathrm{eff}}\right|_{\mathcal{O}(m^2)}=i\int\dd^dx\frac{\dd^dq}{(2\pi)^d}&\int^m \dd m'\,\tr\,\Big( \Delta Q_o\Delta Q_o\Delta + \Delta Q_e\Delta Q_e\Delta\\
&+\Delta Q_o\Delta (-i\sdc)\Delta+\Delta (-i\sdc)\Delta Q_o\Delta + \Delta(-i\sdc)\Delta(-i\sdc)\Delta\Big)\, .
\end{split}
\end{align}
Recall that if the fermion is chiral, we choose to put the axial field $\slashed A\gamma_5$ in $Q_o$ and keep $(\dc_\mu\psi)=(\pa_\mu+iV_\mu+\omega_\mu)\psi$.
We follow the systematic procedure from Section~\ref{SystematicProcedure}, taking care of the spin-connection from the open derivatives to the right. In RNC we obtain directly the covariant form,
\begin{align}
\begin{split}
\left.\Lagr^{\mathrm{ferm}}_{\mathrm{eff}}\right|_{\mathcal{O}(m^2)}=\g\frac{-1}{16\pi^2}m^2\tr\Bigg[ &4\left(1-\lm\right)A^\mu A_\mu+2\left(1-3\lm\right)W_0^2\\
&+2\left( 3-\lm \right)W_1^2-\frac{1}{6}R\left( 1-\lm \right) \Bigg]\, .
\end{split}
\end{align}
The remaining trace is over gauge degrees of freedom.
We remark that no cross term between curvatures and $Q$ appears (no covariant operator can be written).

We ensured of the covariance of the result by conducting the expansion without RNC, and then forming the curvature invariants by collecting the Christoffel symbols and their derivatives together.\\

At order $m$ we have the following terms,
\begin{align}
\begin{split}
\left.S^{\mathrm{ferm}}_{\mathrm{eff}}\right|_{\mathcal{O}(m)}=i\int\dd^dx\frac{\dd^dq}{(2\pi)^d}&\int^m \dd m'\,\tr\, \left[\Delta\left( -i\sdc +Q \right)\right]^3 \Delta\, .
\end{split}
\end{align}
At this order, there is no pure gravity terms because $\left(\Delta(-i\sD)\right)^3\Delta$ vanishes under the Dirac trace.
After following the systematic procedure, we obtain terms involving only $W_0$, $W_1$, $X_\mu$ and $A_\mu$. They agree with the result from \cite{Ellis:2020ivx}, therefore we do not write them down.
The only cross-term between gravity and $Q$ that we obtain is,
\begin{equation}
\left.\Lagr^{\mathrm{ferm}}_{\mathrm{eff}}\right|_{\mathcal{O}(m)}\supset\g\frac{-1}{16\pi^2}m\,\frac{1}{3}\lm\, \tr\,R\,W_0\, .    
\end{equation}
This contribution was already known from previous one-loop computations in curved spacetime since the bosonisation can be performed with $Q=W_0$.
Once again, the computation was performed without the use of RNC to ensure of the covariance of the result.\\

We consider now the terms of order $m^0$. We apply the systematic procedure from Section~\ref{SystematicProcedure}. We omit again the terms that do not involve curvature, they agree with those obtained in \cite{Ellis:2020ivx}. After using Bianchi identities we get,
\begin{align}
\label{m0Ferm}
&\left.\Lagr^{\mathrm{ferm}}_{\mathrm{eff}}\right|_{\mathcal{O}(m^0)}\nonumber\\
&\supset\g\frac{-1}{16\pi^2}\tr\,\Bigg[\lm\left( -\frac{1}{144}R^2+\frac{1}{90}R_{\mu\nu}R^{\mu\nu}+\frac{7}{720}R_{\mu\nu\rho\sigma}R^{\mu\nu\rho\sigma} +\frac{1}{60}(\Box R) \right)\\
&\hspace{1,7cm}+\frac{1}{3}R\, W_0^2\left( 1+\frac{1}{2}\lm \right)+\frac{1}{3}R\, W_1^2\left( -1+\frac{1}{2}\lm \right) -\frac{2}{3}A^\mu A^\nu R_{\mu\nu}\lm\Bigg]\nonumber\, ,
\end{align}
where again the remaining trace is over gauge degrees of freedom.

The pure curvature terms from the first line agree with those that arise after bosonising a vector-like fermion. For example they can be obtained from Eq.~\eqref{bosm0}, if we include the $-\frac{R}{4}\mathds{1}_{\mathrm{Dirac}}\subset U$ and the spin-connection from the field strength \footnote{The pure curvature terms of the previous orders can be obtained from a bosonised vector-like fermion as well.}.
However, the cross-terms between gravity and $W_1$ or $A$ have not been computed before in any functional method.
\\

The first contribution to non-renormalisable operators occurs at the order $1/m$. We only display the terms that depend on the curvature to keep to expression compact. After integrations by parts and use of Bianchi identities we obtain,
\begin{align}
\begin{split}
&\left.\Lagr^{\mathrm{ferm}}_{\mathrm{eff}}\right|_{\mathcal{O}(1/m)}\\
&\supset \g\frac{-1}{16\pi^2}\frac{1}{m}\tr\,\Bigg(W_1\,\left(-\frac{1}{48}\varepsilon_{\mu\nu\rho\sigma}\tensor{R}{_\alpha_\beta^\mu^\nu}\tensor{R}{^\alpha^\beta^\rho^\sigma}\right)\\
&\quad\quad\quad\quad\quad\quad +W_0\left(\frac{1}{45}R^{\mu\nu}R_{\mu\nu}-\frac{1}{72}R^2 +\frac{7}{360}R^{\mu\nu\rho\sigma}R_{\mu\nu\rho\sigma}+\frac{1}{3}(\Box R)\right)\\
&\quad\quad\quad\quad\quad\quad +R^{\mu\nu}\left( -\frac{4}{3}W_0 A_\mu A_\nu+i\frac{4}{3}A_\mu (\dc_\nu W_1)+2iW_1(\dc_\mu A_\nu))+\frac{2}{3}A_\mu [X_\nu,W_1] \right)\\
&\quad\quad\quad\quad\quad\quad+R\left(-\frac{1}{3}A^\mu [X_\mu,W_1]+\frac{1}{9}W_0^3+\frac{1}{3}W_0 W_1^2-i\frac{1}{3}W_1(\dc_\mu A^\mu)\right)\Bigg)\, ,
\label{1/mFerm}
\end{split}
\end{align}
where $\varepsilon_{\mu\nu\rho\sigma}=\g \bar\varepsilon_{\mu\nu\rho\sigma}$ with 
$\bar\varepsilon_{\mu\nu\rho\sigma}$ the Levi-Civita tensor in flat spacetime. The first line of Eq.~\eqref{1/mFerm} corresponds to the axial-gravitational anomaly if we take $W_1= 2\theta m$ arising from an axial field reparametrisation with parameter $\theta$ of the heavy fermion (see Section~\ref{GravAn}). Similarly, the second line corresponds to the Weyl anomaly. The last two lines correspond to new operators that were not computed before (except for the $R\, W_0^3$ term). 

Note that the covariant derivative bears the gauge vector field $V$, and $X$ is a generic vector field. It means that $X$ can also be a gauge field $X=V'$, and in that case it seems that in the last two lines of Eq.~\eqref{1/mFerm} the vector gauge symmetry associated to $V'$ is broken. In fact, this is not the case. If from the start we have a Lagrangian with two vector gauge fields such that $\bar\psi(i\slashed\pa-\slashed V-\slashed V')\psi$, we can choose to proceed with the expansion by keeping $i\sdc\supset -\slashed V$ and deal with the other gauge field with $X=V'$ (recall that in Eq.~\eqref{FermDet} $Q$ comes in with a minus sign already). Using $i(\dc \mathcal{O})=i(\nabla \mathcal{O})-[V,\mathcal{O}]+i[\omega,\mathcal{O}]$ for any operator $\mathcal{O}$, and the cyclicity of the gauge trace, one can show that the last two lines of Eq.~\eqref{1/mFerm} can be rewritten without the explicit $X=V'$ terms and replacing $V$ by $V+V'$ in $i(\dc W_1)$ and $i(\dc A)$. Therefore, the vector gauge symmetry is preserved. \\

Finally, the universal fermionic one-loop effective action in curved spacetime, up to dimension 5 operators, and including only curvature dependent operators, reads,
\begin{align}
\begin{split}
S^{\mathrm{ferm}}_{\mathrm{eff}}\supset\frac{-1}{16\pi^2}&\int\g\dd^4 x\, \tr\,\Bigg\{-m^2\frac{1}{6}R\left( 1-\lm \right)+m\,\frac{1}{3}\lm\, \,R\,W_0\\
&\\
&+\lm\bigg[ -\frac{1}{144}R^2+\frac{1}{90}R_{\mu\nu}R^{\mu\nu}+\frac{7}{720}R_{\mu\nu\rho\sigma}R^{\mu\nu\rho\sigma} +\frac{1}{60}(\Box R)\bigg] \\
&\hspace{2,45cm}+\frac{1}{3}R\, W_0^2\left( 1+\frac{1}{2}\lm \right)+\frac{1}{3}R\, W_1^2\left( -1+\frac{1}{2}\lm \right)\\
&\hspace{2,45cm}-\frac{2}{3}A^\mu A^\nu R_{\mu\nu}\lm \\
&\\
&+\frac{1}{m}\bigg[W_1\,\left(-\frac{1}{48}\varepsilon_{\mu\nu\rho\sigma}\tensor{R}{_\alpha_\beta^\mu^\nu}\tensor{R}{^\alpha^\beta^\rho^\sigma}\right)\\
&\hspace{1cm}+W_0\left(\frac{1}{45}R^{\mu\nu}R_{\mu\nu}-\frac{1}{72}R^2 +\frac{7}{360}R^{\mu\nu\rho\sigma}R_{\mu\nu\rho\sigma}+\frac{1}{3}(\Box R)\right)\\
&\hspace{1cm} +R^{\mu\nu}\left( -\frac{4}{3}W_0 A_\mu A_\nu+i\frac{4}{3}A_\mu (\dc_\nu W_1)+2iW_1(\dc_\mu A_\nu))+\frac{2}{3}A_\mu [X_\nu,W_1] \right)\\
&\hspace{1cm}+R\left(-\frac{1}{3}A^\mu [X_\mu,W_1]+\frac{1}{9}W_0^3+\frac{1}{3}W_0 W_1^2-i\frac{1}{3}W_1(\dc_\mu A^\mu)\right)\bigg]\\
&+\mathcal{O}(1/m^2)\Bigg\}\, .
\label{UOLEAgravFerm}
\end{split}
\end{align}
The terms that involve only $W_0$ and curvatures invariants can be recovered from Eq.~\eqref{UOLEAgravBos} by bosonising the functional determinant involving vector-like fermions, they are in agreement with \cite{Avramidi:2000bm}. The rest consist in new operators and involve the curvature invariants and the fields $A$ and $W_1$, that chiraly couple to the integrated out fermion. To the best of our knowledge, these operators are new and were never computed before in the path integral approach.

\subsection{An example: axial-gravitational anomaly}
\label{GravAn}

These results which have been derived and their associated procedure are quite powerful and we give another example in the context of QFT anomalies. Indeed, the CDE in curved spacetime can be used to compute gravitational anomalies. We follow \cite{Filoche:2022dxl} for the derivation of the Jacobian~\footnote{Another approach that relies on the insertion of a free parameter in the Fujikawa regulator can be found in \cite{Urrutia:1992mx,Cohen:2023hmq}}. Let's illustrate this by computing the axial-gravitational anomaly.
Under an axial reparametrisation of the fermions $\psi\to e^{i\theta\gamma_5}\psi$ and $\bar\psi\to\bar\psi e^{i\theta\gamma_5}$, the path integral measure transforms with a non-trivial Jacobian which can be written as,
\begin{equation}
J[\theta]=\frac{\det\,\g\left(i\sdc-m\right)}{\det\,\g\left(i\sdc-m-2im\theta\gamma_5-(\slashed\pa\theta)\gamma_5\right)}\, .
\end{equation}
The Jacobian is therefore expressed as a ratio of effective field theories and, for $\theta$ infinitesimal, it can be written as,
\begin{align}
\begin{split}
\log J[\theta]&=-\Tr\log\,\g\left.\left( i\sdc-2im\theta\gamma_5-(\slashed\pa\theta)\gamma_5\right)\right|_{\mathcal{O}(\theta)}\\
&=-\left.S^{\mathrm{ferm}}_{\mathrm{eff}}\right|_{\mathcal{O}(\theta)}\, ,
\end{split}
\end{align}
where the subscript $\mathcal{O}(\theta)$ means that only the terms linear in $\theta$ contribute.

The question of hermiticity of the Dirac operator and the choice of scheme for $\gamma_5$ are crucial in the computation of anomalies. More details can be found in \cite{Filoche:2022dxl}.

The only contribution to the Jacobian is at order $m^0$,
\begin{equation}
\log J[\theta] = \int\dd^dx\dq \int\dd m'\,\left.\tr\left(\,\left[(-i\sdc+2im\theta\gamma_5)\right]^5\Delta+\left[\Delta(-i\sdc+(\slashed\pa\theta)\gamma_5)\right]^4\Delta\right)\right|_{\mathcal{O}(\theta)}\, .
\end{equation}
The contribution from $2im\theta\gamma_5$ is already computed in the term of order $1/m$ taking $iW_1=2im\theta$ and according to Eq.~\eqref{1/mFerm} we have,
\begin{equation}
\int\dd^dx\dq \int\dd m'\,\left.\tr\,\left[(-i\sdc+2im\theta\gamma_5)\right]^4\Delta\right|_{\mathcal{O}(\theta)}=\frac{-i\theta}{384\pi^2} \varepsilon_{\mu\nu\rho\sigma}\tensor{R}{_\alpha_\beta^\mu^\nu}\tensor{R}{^\alpha^\beta^\rho^\sigma}\, .
\end{equation}
Note that this contribution involves only finite integrals, thus the computation is performed in $d=4$ dimensions.

The contribution from $(\slashed\pa\theta)\gamma_5$ is also already computed in the term of order $m^0$ if we take $iA_\mu=(\pa_\mu\theta)$. Note that it is a divergent contribution, which thus depends on the $\gamma_5$ scheme. A careful treatment, without using the RNC, reveals that it vanishes \footnote{The derivative term can contribute for example in non-abelian anomalies. Although it is divergent, the result is finite after regularisation. In other words, the pole $2/\epsilon$ cancels. }.

Finally we obtain the axial-gravitational anomaly,
\begin{equation}
\log J[\theta]=\theta\frac{-i}{384\pi^2} \varepsilon_{\mu\nu\rho\sigma}\tensor{R}{_\alpha_\beta^\mu^\nu}\tensor{R}{^\alpha^\beta^\rho^\sigma}\, ,
\end{equation}
which corresponds to the well-known result \cite{Bertlmann:1996xk,Fujikawa:2004cx}.

\section{Conclusion}

We presented a new method for deriving loop corrections in curved spacetime within the path integral, based on the Covariant Derivative Expansion (CDE), with a coordinate independent momentum representation. The proposed procedure is less computationally involved compared to previous approaches. It allows to derive for the first time curved spacetime EFTs incorporating non-renormalisable operators within the CDE approach. This extends the so-called Universal One Loop Effective Action (UOLEA) so far available in flat spacetime to curved spacetime. Our results are summarised in Eqs.~\eqref{UOLEAgravBos} and \eqref{UOLEAgravFerm} and correspond to gravity induced operators for the bosonic UOLEA and the fermionic UOLEA, up to respectively dimension six and dimension five operators.

A significant novelty of the approach is the integration of a chiral fermion in curved spacetime which was never performed before within the functional approach. This generates new operators in the EFT, both renormalisable and non-renormalisable, that can be found in Eq.~\eqref{UOLEAgravFerm}. To the best of our knowledge, this offers the only alternative to the use of Feynman diagrams when dealing with a chiral fermion in curved spacetime, with the advantage of not requiring an expansion in the metric.

In addition to these effective actions, one significant outcome of this work, is the systematic procedure which allows to straightforwardly evaluate even higher dimensional operators. Each term is computed following the same systematic procedure: commute the momentum dependence to the left of the covariant derivatives using Table~\ref{tab1}, in order to perform the integration in terms of master integrals, then form covariant quantities, which can be greatly simplified using the Riemann Normal Coordinates and the Fock-Schwinger gauge.

This method would very well fit in a code that performs CDE to achieve one-loop matching (see for example \cite{Cohen:2020qvb,Fuentes-Martin:2020udw,Fuentes-Martin:2022jrf}). It would allow to integrate out fields on a curved spacetime background at one-loop, and even spin-2 fields (for example the graviton) at one-loop, which is not implemented so far in EFT matching codes \cite{Cohen:2020qvb,Fuentes-Martin:2020udw,Carmona:2021xtq,Fuentes-Martin:2022jrf}.

The transparency of the expansion offers physical insight into the computation, which is rather concealed under mathematical complexity in other approaches such as heat kernel in position space and other attempts in momentum space.

The recent developments on the CDE technique which enable the derivation of EFTs induced by loops of fields with non-degenerate mass, mixed heavy-light particles, and more, can be easily combined with this new expansion in curved spacetime.

Our results could have multiple applications and we gave an example of how it can be used to efficiently compute anomalies.
The computational transparency and simplicity of the presented method makes it a powerful tool to study for example inflation or low energy effects of a UV completion of gravity.

\section*{Acknowledgments} 
The authors are grateful to Pham Ngoc Hoa Vuong for useful discussions, Rodrigo Alonso, Philippe Brax, Baptiste Filoche and Tevong You for helpful comments on the manuscript. This work is supported by the IN2P3 Master projects A2I and BSMGA and by the Programme National GRAM of CNRS/INSU with INP and IN2P3 co-funded by CNES. R.L. acknowledges the support of the European Consortium for Astroparticle Theory in the form of an Exchange Travel Grant.

\appendix

\section{Master integrals}
\label{Appendix:master-integrals}

\subsection*{Master integrals in curved spacetime}

In flat spacetime (latin indices), we define the master integrals $\mathcal{I}$, $\mathcal{J}$ and $\mathcal{K}$ as,
\begin{align}
&\int \dq \dfrac{q^{a_1}\cdots q^{a_{2n_c}}}{(q^2-m^2)^n }= \eta^{a_1 \cdots a_{2n_c}} \mathcal{I}[q^{2n_c}]^n \\
&\int \dq\int^{m^2}\dd m'^2 \dfrac{q^{a_1}\cdots q^{a_{2n_c}}}{(q^2-m'^2)^n }=\eta^{a_1 \cdots a_{2n_c}} \mathcal{J}[q^{2n_c}]^{n}\\
&\int\dq\int^m\dd m'\,m'^k\frac{q^{a_1}\dots q^{a_{2n_c}}}{(q^2-m'^2)^n}=\eta^{a_1\dots a_{2n_c}}\mathcal{K}[q^{2n_c}]^k_n\, ,
\end{align}
where in general the integral over the mass must be performed before the integral over momentum. $\eta^{a_1\dots a_{2n_c}}$ is the fully symmetrised Minkowski metric.

The master integrals, $\mathcal{I}$ are defined by the general expression,
\begin{equation}
\mathcal{I}[q^{2n_c}]^{n} = \frac{i}{16\pi^2} \bigl(-m^2\bigr)^{2+n_c-n}
\frac{1}{2^{n_c}(n-1)!} \frac{\Gamma(\frac{\epsilon}{2}-2-n_c +n)}{\Gamma(\frac{\epsilon}{2})} \Bigl(\frac{2}{\epsilon} -\gamma +\log 4\pi-\log\frac{m^2}{\mu^2}\Bigr) \,,
\label{masterI}
\end{equation}
where $d=4-\epsilon$ is the spacetime dimension, and $\mu$ is the renormalization scale. In the $\overline{\rm MS}$ scheme, we replace, $\bigg(\dfrac{2}{\epsilon} -\gamma \, + \, \log 4\pi -\log\dfrac{m^2}{\mu^2}\bigg)$ by $\bigg(-\log\dfrac{m^2}{\mu^2}\bigg)$ in the final result. We factor out the common prefactor, $\mathcal{I}=\frac{i}{16\pi^2}\tilde{\mathcal{I}}$ and present a table of $\tilde{\mathcal{I}}[q^{2n_c}]^{n}$ for various $n_c$ and $n$, needed in our computations, in Table~\ref{tab:MIheavy}.

$\mathcal{I}$ and $\mathcal{J}$ are related by integrating the mass,
\begin{equation} \mathcal{J}[q^{2n_c}]^{n}=\frac{1}{n-1}\mathcal{I}[q^{2n_c}]^{n-1}\, .
\end{equation}

The fermionic master integrals $\mathcal{K}$ are trickier to compute. The dimensionful integrals are computed in Eq.~\eqref{masterKdimensionful}. The dimensionless integrals (i.e $\propto m^0$) can however be obtained using Eqs.~\eqref{masterKiterative} and Eq.~\eqref{masterKiterative2}.

\begin{table}[htbp!]
\centering
\begin{tabular}{|c|ccc|}
\hline
$\tilde{\mathcal{I}}[q^{2n_c}]^{n}$ & $n_c=0$ & $n_c=1$ & $n_c=2$ \T\B  \\
\hline
$n=1$ \T
& $m^2 \bigl(1-\logm{m^2}\bigr)$ 
& $\frac{m^4}{4} \bigl(\frac{3}{2} -\logm{m^2}\bigr)$ 
& $\frac{m^6}{24} \bigl(\frac{11}{6} -\logm{m^2}\bigr)$ \\
$n=2$ 
& $-\logm{m^2}$ 
& $\frac{m^2}{2} \bigl(1 -\logm{m^2}\bigr)$ 
& $\frac{m^4}{8} \bigl(\frac{3}{2} -\logm{m^2}\bigr)$ \\
$n=3$ 
& $-\frac{1}{2m^2}$ 
& $-\frac{1}{4}\logm{m^2}$ 
& $\frac{m^2}{8} \bigl(1 -\logm{m^2}\bigr)$ \\
$n=4$ 
& $\frac{1}{6m^4}$ 
& $-\frac{1}{12m^2}$
& $-\frac{1}{24}\logm{m^2}$\\
$n_i=5$
& $-\frac{1}{12m^6}$
& $\frac{1}{48m^4}$
& $-\frac{1}{96m^2}$ \B \\
\hline
\end{tabular}
\caption{Commonly-used master integrals with degenerate heavy particle masses. $\tilde{\mathcal{I}}=\mathcal{I}/\frac{i}{16\pi^2}$ and the $\frac{2}{\epsilon} -\gamma +\log 4\pi$ contributions are dropped.}
\label{tab:MIheavy}
\end{table}

We can relate the integrals in curved spacetime to those in flat spacetime using a tangent frame that is orthonormal everywhere in the whole manifold. We relate the flat metric $\eta$ and the metric $g$ using the vierbein,
\begin{equation}
g_{\mu\nu}=\tensor{e}{_\mu^a}\tensor{e}{_\nu^b}\eta_{ab}\, .
\end{equation}
The latin indices refer to the orthonormal frame, while the greek indices refer to the initial frame.
The momenta are expressed in the orthonormal frame using the vierbein and its inverse $E$,
\begin{equation}
p_\mu=\tensor{e}{_\mu^a}q_a \quad q_a=\tensor{E}{_a^\mu}p_\mu\, .
\end{equation}
We can now relate the master integrals with momenta $p^\mu$ to the master integrals in flat spacetime by doing the change of variable $p_\mu=\tensor{e}{_\mu^a}q_a$. The momentum space measure is defined with the covariant vector $p_\mu$, thus the jacobian of the change of variable is $\det(e)=\g$. We thus have,
\begin{align}
\begin{split}
&\int \frac{\dd^d p}{(2\pi)^4}\int^{m^2}\dd m'^2 \dfrac{p^{\mu_1}\cdots p^{\mu_{2n_c}}}{(p^2-m'^2)^n }\\
&=\int \det(e)\frac{\dd^d q}{(2\pi)^4}\int^{m^2}\dd m'^2 \tensor{E}{^{\mu_1}_{a_1}}\cdots\tensor{E}{^{\mu_{2n_c}}_{a_{2n_c}}}\dfrac{q^{a_1}\cdots q^{a_{2n_c}}}{(q^2-m'^2)^n }\\
&=\g \tensor{E}{^{\mu_1}_{a_1}}\cdots\tensor{E}{^{\mu_{2n_c}}_{a_{2n_c}}}\eta^{a_1 \cdots a_{2n_c}} \mathcal{J}[q^{2n_c}]^{n}\\
&=\g g^{\mu_1\cdots\mu_{2n_c}}\mathcal{J}[q^{2n_c}]^{n}\, ,
\end{split}
\end{align}
and likewise for $\mathcal{K}$ and $\mathcal{I}$.

\subsection*{Momentum and mass integration}

The master integrals with integration over the mass should be computed by integrating over the mass first, and then over momentum. In general commuting the integration is not true for divergent integrals. However, we will show that it stands true for dimensionful integrals.

The discussion below is not very relevant for the bosonic integrals $\mathcal{J}$ since we are able to perform the integration over the mass and then over momentum, as it should be done. However, it is not so simple for the fermionic integrals, therefore commuting the integrals will prove useful.

Let's reason on the bosonic integrals which are simpler to compute explicitly. Without commuting the integrals, we can perform the integration over the mass to obtain the correct result,
\begin{equation}
\mathcal{J}[q^{2l}]^n=\frac{1}{n-1}\mathcal{I}[q^{2l}]^{n-1}\, .
\label{masterJ1}
\end{equation}
If we commute the integrals, then we obtain for $2+l-(n-1)\neq0$,
\begin{align}
\begin{split}
\mathcal{J}'[q^{2l}]^n&=\int^{m^2}\dd m'^2\mathcal{I}[q^{2l}]^n(m')\\
&=\frac{i}{16\pi^2}\frac{(m^2)^{2+l-(n-1)}}{2+l-(n-1)}\frac{(-1)^{l-n}}{2^l(n-1)!}\frac{\Gamma(\frac{\epsilon}{2}-2-l+n)}{\Gamma(\frac{\epsilon}{2})}\left(\frac{1}{2+l-(n-1)}+\frac{2}{\bar\epsilon}-\lm\right)\, .
\label{masterJ2}
\end{split}
\end{align}
Note that the integration constant vanishes by dimensional analysis since by definition it must be independent of the mass, but the integral is dimensionful for $2+l-(n-1)\neq0$. We defined $2/\bar\epsilon=2/\epsilon-\gamma+\log 4\pi$.

If $(n-1)-l-2>0$, then both $\mathcal{I}[q^{2l}]^{n-1}$ and $\mathcal{J}[q^{2l}]$ are finite in $d=4$ dimensions. The commutation of the integral is thus correct, and we can indeed verify that Eqs.~\eqref{masterJ1} and \eqref{masterJ2} are equal.

However, we can show that the commutation of the integrals remains true if $(n-1)-l-2<0$. Using for $N\geq0$,
\begin{equation}
\frac{\Gamma(\frac{\epsilon}{2}-N)}{\Gamma(\frac{\epsilon}{2})}=\frac{(-1)^N}{N!}\left(1+\frac{\epsilon}{2}\sum_{k=1}^{N-1}1/k+\mathcal{O}(\epsilon^2)\right)\, ,
\end{equation}
and the expression of $\mathcal{I}$ in Eq.~\eqref{masterI}, we can show that Eqs.~\eqref{masterJ1} and \eqref{masterJ2} remain equal.

A discrepancy between $\mathcal{J}[q^{2l}]^n$ and $\mathcal{J}'[q^{2l}]^n$ however happens when $(n-1)-l-2=0$ since we have,
\begin{equation}
\mathcal{J}[q^{2l}]^n=\frac{i}{16\pi^2}\frac{1}{2^l(n-1)!}\left(\frac{2}{\bar\epsilon}-\lm\right)\, ,
\label{m0masterJ1}
\end{equation}
whereas,
\begin{equation}
\mathcal{J}'[q^{2l}]^n=\frac{i}{16\pi^2}\int^{m^2}\dd m'^2\frac{-1}{2^l(n-1)!m'^2}+\mathcal{O}(\epsilon)=\frac{i}{16\pi^2}\frac{-1}{2^l(n-1)!}\lm+F\, ,
\label{m0masterJ2}
\end{equation}
where $F$ is an integration constant that cannot be ruled out by dimensional analysis like before.

Now Eqs.~\eqref{m0masterJ1} and \eqref{m0masterJ2} are not equal. Note however, that the coefficient of the logarithm in Eq.~\eqref{m0masterJ2} is correct, and the difference lies in the undetermined integration constant.\\

As explained earlier, commuting the integrals is not useful for the bosonic integrals since we are able to easily perform the integration over the mass then over momentum. However, some fermionic integrals are much easier to compute if we are allowed to commute the integrals.

The fermionic integrals $\mathcal{K}$ are more cumbersome to compute in general because the integration is over the mass instead of the mass square. For example, after Wick rotation, the integration over the mass of $\mathcal{K}[q^{2l}]^0_1$ yields,
\begin{equation}
\mathcal{K}[q^0]^0_1\propto\int\frac{\dd^dq}{(2\pi)^d}q^{2l}\int^m\dd m'\,\frac{1}{q^2-m'^2}=-i\int\frac{\dd^dq}{(2\pi)^d}q^{2l-1}\mathrm{Arctan}\left(\frac{m}{q}\right)\, ,
\label{KpblExample}
\end{equation}
where $q=\sqrt{q^2}$ is well-defined since $q^2\geq0$ in Euclidian.
The integration over momentum is then not trivial to perform in dimensional regularisation.

However, as shown for the bosonic integrals, the integration over mass and momentum can be commuted provided $\mathcal{K}$ is dimensionful \footnote{We do not prove it for the fermionic integrals, we assume it behaves similarly.}.

Hence, for $k+2l-2n+5\neq0$ (i.e for dimensionful integrals), we have,
\begin{align}
\begin{split}
&\mathcal{K}[q^{2l}]^k_n=\int^m\dd m'\,m'^k \mathcal{I}[q^{2l}]^n\\
&=\frac{i}{16\pi^2}\frac{(-1)^{l-n}m^{5+k+2l-2n}}{2^l(n-1)!(5+k+2l-2n)}\frac{\Gamma(\frac{\epsilon}{2}-2-l+n)}{\Gamma(\frac{\epsilon}{2})}\left(\frac{2}{5+k+2l-2n}+\frac{2}{\bar\epsilon}-\lm\right)\, .
\label{masterKdimensionful}
\end{split}
\end{align}

The dimensionless integrals can be computed using recursion formulae which can be obtained by integration by parts over the mass integration,
\begin{align}
&\mathcal{K}[q^{2l}]^k_n=\frac{m^{k-1}}{2(n-1)}\mathcal{I}[q^{2l}]^{n-1}-\frac{k-1}{2(n-1)}\mathcal{K}[q^{2l}]^{k-2}_{n-1}\label{masterKrec}\\
&\mathcal{K}[q^{2l}]^k_n=\frac{m^{k-1}}{2n-k-1}\mathcal{I}[q^{2l}]^{n-1}-\frac{k-1}{2n-k-1}(d+2l)\mathcal{K}[q^{2(l+1)}]^{k-2}_n\label{masterKrec2}\\
&\mathcal{K}[q^{2l}]^k_{n}=\frac{2n}{2n-k-1}(d+2l)\mathcal{K}[q^{2(l+1)}]^k_{n+1}-\frac{m^{k+1}}{2n-k-1}\mathcal{I}[q^{2l}]^{n}\label{masterKrec3}\, ,
\end{align}
where $d$ is the dimension of spacetime \footnote{It arises due to the scalarisation of the vectors $q$. In the momentum integrals, $q_{\mu_1}\dots q_{\mu_{2l}}$ is traded for $c_l q^{2l}g_{\mu_1\dots\mu_{2l}}$ where $c_l^{-1}=d(d+2)\dots (d+2(l-1))$.}.

Using repeatedly Eq.~\eqref{masterKrec}, we obtain,
\begin{align}
\begin{split}
    \mathcal{K}[q^{2l}]^{2k+1}_n=&\frac{1}{2}\sum_{i=1}^k(-1)^{k-i}\frac{k!}{i!}\frac{(n-2-k+i)!}{(n-1)!}m^{2i}\mathcal{I}[q^{2l}]^{n-1-k+i}\\
    &+(-1)^{k}k!\frac{(n-k-1)!}{(n-1)!}\mathcal{K}[q^{2l}]^{1}_{n-k}\, ,
    \label{masterKiterative}
\end{split}
\end{align}
and for $n-k>1$ \footnote{In our expansion, the integrals $\mathcal{K}[q^{2kl}]^p_n$ are such that $n\geq p$, so if $p=2k+1$, $n-k=1$ can only be realised for $n=1$ which contributes as a tadpole to the effective action.}, we have,
\begin{equation}
\mathcal{K}[q^{2l}]^{1}_{n-k}=\frac{1}{2(n-k-1)}\mathcal{I}[q^{2l}]^{n-k-1}\, .
\label{masterKiterative2}
\end{equation}

Note however, that integrals of the form $\mathcal{K}[q^{2l}]^{2k}_n$ remain troublesome to compute even using the iterative formulae. For example, using Eqs.~\eqref{masterKrec2} and \eqref{masterKrec3} repeatedly, $\mathcal{K}[q^{2l}]^{2k}_n$ can be related to $\mathcal{K}[q^{2l'}]^{0}_1$ from Eq.~\eqref{KpblExample}, which is not trivial to compute.

The dimensionless integrals only occur in our expansion in the term,
\begin{equation}
\int\frac{\dd^dq}{(2\pi)^d}\int^m\dd m' \left[\Delta(-i\sdc+Q)\right]^4\Delta\, .
\end{equation}
Since the number of propagators is odd, and the power in $q$ in the numerator must be even lest the integral vanishes, then the power in $m'$ in the numerator is odd. Therefore, these integrals can be computed using Eqs.~\eqref{masterKiterative} and \eqref{masterKiterative2}.

\section{Momentum representation in curved spacetime}
\label{Appendix:MomentumManifold}

In this Appendix we present the difficulties that arise with the momentum representation in curved spacetime, and how to define it in a coordinate independent manner.

\subsection*{Momentum representation}

We consider a manifold $\mathcal{M}$ of dimension $d$, provided with an atlas $\{(U_i,\phi_i)\}$. 
Let $p$ be a point in $\mathcal{M}$. There exists a chart $(U,\phi)$ in the atlas such that $p\in U$. $\phi(p)\in \mathds{R}^d$ is the coordinate representation of $p$, we note $\phi^\mu(p)=x^\mu(p)$.
We can choose a set of $d$ functions from $\mathds{R}^d\to\mathds{R}$: $q_\mu(x)$, $\mu\in \{1,\dots,d\}$, such that,
\begin{equation}
\frac{\pa q_\mu}{\pa x^\nu}=(\pa_\nu q_\mu)=0\, . 
\end{equation}
This is simply the momentum conjugate to $x$ in the flat space $\mathds{R}^d$.
We can thus provide the points in $U$ with a momentum representation with,
\begin{equation}
e^{-iq\cdot x}\frac{\pa}{\pa x^\mu} e^{iq\cdot x}=\pa_\mu+iq_\mu\, .
\end{equation}

One would be tempted to define the 1-form $Q=q_\mu dx^\mu$ and the vector $X=x^\mu\frac{\pa}{\pa x^\mu}$ so that $q\cdot x$ is coordinate invariant. But $X$ does not define a vector: suppose we have a second chart such that $p\in V$ with a coordinate $\varphi^\mu(p)=y^\mu(p)$, then $x$ and $y$ are related by a diffeomorphism $y^\mu=f^\mu(x)$ and we do not have in general $y^\mu=x^\nu\frac{\pa y^\mu}{\pa x^\nu}$.

The second chart $(V,\varphi)$ provides another momentum representation $r^\mu$ such that $\frac{\pa r^\mu}{\pa y^\nu}=0$. Consider a vector $T=T^\mu\frac{\pa}{\pa x^\mu}=\tilde T^\mu\frac{\pa}{\pa y^\mu}$. 
We can write,
\begin{equation}
e^{-ir.y}Te^{ir.y}=\tilde T^\mu e^{-ir_\nu y^\nu}\frac{\pa}{\pa y^\mu}e^{ir_\nu y^\nu}=\tilde T^\mu\left(\frac{\pa}{\pa y^\mu}+ir_\mu\right)\, .
\end{equation}
On the other hand,
\begin{align}
\begin{split}
e^{-ir.y}Te^{ir.y}&=T^\mu e^{-ir_\nu f^\nu(x)}\frac{\pa}{\pa x^\mu}e^{ir_\nu f^\nu(x)}\\
&=T^\mu\left(\frac{\pa}{\pa x^\mu}+i\frac{\pa r_\nu}{\pa x^\mu}f^\nu(x)+ir_\nu\frac{\pa f^\nu(x)}{\pa x^\mu}\right)\\
&=T^\mu\left(\frac{\pa}{\pa x^\mu}+ir_\nu\frac{\pa f^\nu(x)}{\pa x^\mu}\right)\, ,
\end{split}
\end{align}
where we used,
\begin{equation}
\frac{\pa r_\nu}{\pa x^\mu}=\frac{\pa y^\rho}{\pa x^\mu}\frac{\pa r_\nu}{\pa y^\rho}=0\, .
\end{equation}
Since we defined the vector $Q=r_\mu dy^\mu=q_\mu dx^\mu$, $q_\mu$ transforms covariantly: $q_\mu=r_\nu \frac{\pa y^\nu}{\pa x^\mu}=r_\nu \frac{\pa f^\nu(x)}{\pa x^\mu}$. Hence,
\begin{equation}
e^{-ir.y}Te^{ir.y}=T^\mu\left(\frac{\pa}{\pa x^\mu}+iq_\mu\right)=e^{-iq\cdot x} T e^{iq\cdot x}\, .
\end{equation}

In summary, if we have two coordinates $\{x^\mu\}$ and $\{y^\mu\}$ for a given point $p\in\mathcal{M}$, we can define the vector $Q=q_\mu dx^\mu=r_\mu dy^\mu$ such that $\frac{\pa q_\mu}{\pa x^\nu}=\frac{\pa r_\mu}{\pa y^\nu}=0$, and we have for a vector $T=T^\mu \frac{\pa}{\pa x^\mu}=\tilde T^\mu\frac{\pa }{\pa y^\mu}$,
\begin{align}
\begin{split}
&e^{-iq\cdot x} T e^{iq\cdot x}=T^\mu(\frac{\pa}{\pa x^\mu}+iq_\mu)\\
=&e^{-ir\cdot y}  T e^{ir\cdot y}=\tilde T^\mu(\frac{\pa}{\pa y^\mu}+ir_\mu)\, .
\end{split}
\end{align}
Note however that $q\cdot x\neq r\cdot y$,
\begin{equation}
q\cdot x=q_\mu x^\mu=r_\nu\frac{\pa y^\nu}{\pa x^\mu}x^\mu\neq r_\nu y^\nu\, .
\end{equation}
Besides since $y=f(x)$, we have,
\begin{equation}
\dd^d y = \dd y^1 \wedge\dots\wedge\dd y^d=\dd(f^1(x))\wedge\dots\wedge\dd(f^d(x))=\frac{\pa f^1(x)}{\pa x^{\mu_1}}\dd x^{\mu_1}\wedge\dots\wedge \frac{\pa f^d(x)}{\pa x^{\mu_d}}\dd x^{\mu_d}=\det\left(\frac{\pa f(x)}{\pa x}\right)\dd^dx\, ,
\end{equation}
and,
\begin{equation}
\dd^d q=\dd(\frac{\pa f^{\mu_1}(x)}{\pa x^{1}}r_{\mu_1})\wedge\dots\wedge\dd(\frac{\pa f^{\mu_d}(x)}{\pa x^{d}}r_{\mu_d})=\det\left(\frac{\pa f(x)}{\pa x}\right)\dd^d r\, .
\end{equation}
We thus have the invariance of the measure,
\begin{equation}
\dd^d x\dd^d q=\dd^d y \dd^d r\, .
\end{equation}

Finally, consider a vector $T$ and the two charts $(U,\phi)$ and $(V,\varphi)$with respective coordinate and momentum $(x,q)$ and $(y,r)$. If they have an intersection $V\cap U\neq \emptyset$, we have,
\begin{equation}
\int_{U\cap V}\dd^dx\frac{\dd^d q}{(2\pi)^d}e^{-iq\cdot x}T e^{iq\cdot x}=\int_{U\cap V}\dd^dy\frac{\dd^d r}{(2\pi)^d}e^{-ir\cdot y}T e^{ir\cdot y}\, .
\end{equation}

This result can be generalised without difficulty to any tensor $H=H^{\mu_1\dots\mu_n}\pa_{\mu_1}\dots\pa_{\mu_n}$ by inserting $1=e^{-iq\cdot x}e^{iq\cdot x}$ between each derivative and using the transformation of a rank $n$ tensor under a diffeomorphism.

\subsection*{Functional trace}

We now seek to write the functional trace of the logarithm of an operator $\mathcal{O}$ that is covariant. $\mathcal{O}$ is quadratic in covariant derivatives, it can be written under the form $\mathcal{O}=A^{\mu\nu}\pa_\mu\pa_\nu + B^\mu \pa_\mu + C$ where $A$, $B$ and $C$ depend on the connections in the covariant derivative. The logarithm can be expanded in a serie,
\begin{equation}
\log\mathcal{O}=\sum_{n\geq0}T_n^{\mu_1\dots\mu_n}\pa_{\mu_1}\dots\pa_{\mu_n}\, .
\end{equation}
Since $\mathcal{O}$ is covariant, $T_n$ must be a rank $n$ tensor. According to the previous section,
\begin{equation}
e^{-iq\cdot x}\log\mathcal{O}(x,i\pa_x)e^{iq\cdot x}=\sum_{n\geq0}T_n^{\mu_1\dots\mu_n}(\pa_{\mu_1}+iq_{\mu_1})\dots(\pa_{\mu_n}+iq_{\mu_n})=\log\mathcal{O}(x,i\pa_x-q)\, ,
\end{equation}
is coordinate independent.

By choosing a chart at each point of the manifold, we can write the functional trace,
\begin{equation}
\Tr\,\mathcal{O}=\int_{p\in\mathcal{M}}\dd^d x_i(p)\frac{\dd^d q_i}{(2\pi)^d}\tr\,\mathcal{O}(x_i(p),i\pa_{x_i}-q_i)\, ,
\end{equation}
where $i$ refers to a chart $(U_i,\phi_i)$ such that $p\in U_i$, $x_i(p)=\phi_i(p)$ and $q_i$ is the associated momentum. As shown above, the integration is independent of the choice of chart (i.e coordinate) for each point $p$, which allows us to define the functional trace on a generic manifold $\mathcal{M}$,
\begin{equation}
\Tr\,\mathcal{O}=\int_{p\in\mathcal{M}}\dd^d x(p)\frac{\dd^d q}{(2\pi)^d}\tr\,\mathcal{O}(x(p),i\pa_{x}-q)\, .
\end{equation}

\subsection*{$\pa_\mu q_\nu=0$ versus $\pa_\mu q^\nu=0$}

We have chosen to define the vector $Q$ such that $\frac{\pa q_\mu}{\pa x^\nu}=0$. But we could have chosen $\frac{\pa q^\mu}{\pa x^\nu}=0$, which is not equivalent.
Suppose we make the second choice: $\frac{\pa q^\mu}{\pa x^\nu}=0$. We thus have,
\begin{align}
\begin{split}
e^{-iq\cdot x}T e^{iq\cdot x}&=T^\mu (\frac{\pa}{\pa x^\mu}+iq^\nu\frac{\pa x_\nu}{\pa x^\mu})\\
&=T^\mu (\frac{\pa}{\pa x^\mu}+iq^\nu\frac{\pa g_{\nu\rho}x^\rho}{\pa x^\mu})\\
&=T^\mu (\frac{\pa}{\pa x^\mu}+iq^\nu\frac{\pa g_{\nu\rho}}{\pa x^\mu}x^\rho+iq_\mu)\, ,
\end{split}
\end{align}
which does not yield the desired outcome.

\section{Riemann Normal Coordinates and Fock-Schwinger gauge}
\label{Appendix:RNCFS}

In this Appendix, we provide the expansion of the metric, the Christoffel symbols and the spin-connection in RNC, as well as the expansion of the gauge fields in FS gauge, around $x_0$, with $x=x_0+y$.

\paragraph{Fock-Schwinger gauge}

In the FS gauge around $x_0$, the gauge fields are expressed as follows,
\begin{equation}
V_\mu(x_0)=-\sum_{n\geq0}\frac{1}{n!(n+2)}y^\nu y^{\alpha_1}\dots y^{\alpha_n}\Bigg[D_{\alpha_1}, \bigg[ D_{\alpha_2},\Big[\dots [D_{\alpha_n},F_{\mu\nu}]\dots\Big]\bigg] \Bigg](x_0)\, .
\end{equation}

We must then upgrade the covariant derivatives $D$ to the general covariant derivatives $\dc$ to have a diffeomorphism covariant expression. For example, in RNC and FS gauge we have,
\begin{align}
\begin{split}
[D_\mu,[D_\nu,F_{\rho\sigma}]]y^\mu y^\nu y^\sigma&=[D_\mu,\left([\dc_\nu,F_{\rho\sigma}]+\Gamma^\lambda_{\nu\rho} F_{\lambda\sigma}+\Gamma^\lambda_{\nu\sigma} F_{\nu\lambda}\right)]y^\mu y^\nu y^\sigma\\
&=\left([\dc_\mu,[\dc_\nu,F_{\rho\sigma}]]+(\pa_\mu \Gamma^\lambda_{\nu\rho}) F_{\lambda\sigma}+(\pa_\mu \Gamma^\lambda_{\nu\sigma}) F_{\rho\lambda}\right)y^\mu y^\nu y^\sigma\\
&=\left( [\dc_\mu,[\dc_\nu,F_{\rho\sigma}]] + \frac{1}{3}\tensor{R}{_\mu_\rho^\lambda_\nu}\tensor{F}{_\lambda_\sigma} \right)y^\mu y^\nu y^\sigma\, .
\end{split}
\end{align}
We thus obtain for the first orders the combination of FS gauge and RNC as in \cite{Alvarez-Gaume:1981exa,Dilkes:1995cu},
\begin{equation}
V_\mu=-\frac{1}{2}F_{\mu\nu}(x_0)\,y^\nu-\frac{1}{3}(\dc_\alpha F_{\mu\nu})(x_0)\,y^\nu y^\alpha-\frac{1}{8}\left( (\dc_{\alpha\beta} F_{\mu\nu})+\frac{1}{3}\tensor{R}{_\alpha_\mu^\lambda_\beta}F_{\lambda\nu} \right)(x_0) \,y^\alpha y^\beta y^\nu+\mathcal{O}(y^4)\, .
\label{VFSgauge}
\end{equation}

We do not consider the case of an axial gauge field in the covariant derivative as throughout the computation we put the axial gauge field in $Q_o$ rather than in $\dc$. More crucially, we expect the axial gauge invariance to be broken depending on the choice of scheme for $\gamma_5$ in dimensional regularisation, therefore it makes no sense to choose a gauge for the axial gauge field.

\paragraph{Riemann Normal Coordinates}
The metric and the Christoffel symbols are expressed around $x_0$ as,
\begin{align}
\begin{split}
g_{\mu\nu}(x)=&\eta_{\mu\nu}-\frac{1}{3}\tensor{R}{_\mu_\alpha_\nu_\beta}(x_0)\,y^\alpha y^\beta-\frac{1}{6}\tensor{R}{_\mu_\alpha_\nu_\beta_;_\gamma}(x_0)\,y^\alpha y^\beta y^\gamma\\
&+\left(-\frac{1}{20}\tensor{R}{_\mu_\alpha_\nu_\beta_;_\gamma_\delta}+\frac{2}{45}\tensor{R}{_\alpha_\mu_\beta_\lambda}\tensor{R}{^\lambda_\gamma_\nu_\delta}\right)(x_0)\,y^\alpha y^\beta y^\gamma y^\delta +\mathcal{O}(y^5)\, ,
\end{split}
\end{align}
\begin{align}
\begin{split}
g^{\mu\nu}(x)=&\eta^{\mu\nu}+\frac{1}{3}\tensor{R}{^\mu_\alpha^\nu_\beta}(x_0)\,y^\alpha y^\beta+\frac{1}{6}\tensor{R}{^\mu_\alpha^\nu_\beta_;_\gamma}(x_0)\,y^\alpha y^\beta y^\gamma\\
&+\left(\frac{1}{20}\tensor{R}{^\mu_\alpha^\nu_\beta_;_\gamma_\delta}+\frac{1}{15}\tensor{R}{_\alpha^\mu_\beta_\lambda}\tensor{R}{^\lambda_\gamma^\nu_\delta}\right)(x_0)\,y^\alpha y^\beta y^\gamma y^\delta +\mathcal{O}(y^5)\, ,
\end{split}
\end{align}
and,
\begin{align}
\begin{split}
\Gamma^\mu_{\nu\rho}=&-\frac{1}{3}\left(\tensor{R}{^\mu_\nu_\rho_\alpha}+\tensor{R}{^\mu_\rho_\nu_\alpha}\right)(x_0)\,y^\alpha\\
&-\frac{1}{12}\left( 2\tensor{R}{^\mu_\nu_\rho_\alpha_;_\beta}+2\tensor{R}{^\mu_\rho_\nu_\alpha_;_\beta}+\tensor{R}{^\mu_\alpha_\rho_\beta_;_\nu}+\tensor{R}{^\mu_\alpha_\nu_\beta_;_\rho}-\tensor{R}{_\nu_\alpha_\rho_\beta_;^\mu} \right)(x_0)\,y^\alpha y^\beta\\
&+\Bigg[ \frac{1}{18}\tensor{R}{^\mu_\alpha^\lambda_\beta}y^\alpha y^\beta\left( 
-R_{\rho\gamma\lambda\delta}(\pa_\nu y^\gamma y^\delta)-R_{\nu\gamma\lambda\delta}(\pa_\rho y^\gamma y^\delta) +R_{\nu\gamma\rho\delta}(\pa_\lambda y^\gamma y^\delta)\right)\\
& \quad\quad+ \left.\left(\frac{1}{45}\tensor{R}{_\alpha_\rho_\beta_\lambda}\tensor{R}{^\lambda_\gamma^\mu_\delta}-\frac{1}{40}\tensor{R}{_\alpha_\rho_\beta^\mu_{;\gamma}_\delta} \right)(\pa_\nu y^\alpha y^\beta y^\gamma y^\delta)\right|_{\text{sym }\mu\leftrightarrow\nu}\\
&\quad\quad-\left(\frac{1}{45}\tensor{R}{_\alpha_\nu_\beta_\lambda}\tensor{R}{^\lambda_\gamma_\rho_\delta}-\frac{1}{40}\tensor{R}{_\alpha_\nu_\beta_\rho_{;\gamma}_\delta} \right)g^{\mu\chi}(\pa_\chi y^\alpha y^\beta y^\gamma y^\delta)\Bigg](x_0)+\mathcal{O}(y^4)
\label{ChristoffelRNC}
\end{split}
\end{align}
Higher order expansions in RNC can be found in \cite{Brewin:2009se}.

For example, in the limit $y\to 0$, we have
\begin{align}
\begin{split}
&(\pa_\mu g^{\alpha\beta})=0\\
&(\pa_{\rho\sigma} g^{\mu\nu})=\frac{1}{3}\tensor{R}{^\mu_\alpha^\nu_\beta}\left( \tensor{g}{^\alpha_\rho}\tensor{g}{^\beta_\sigma}+\tensor{g}{^\alpha_\sigma}\tensor{g}{^\beta_\rho} \right)=\frac{1}{3}\left(\tensor{R}{^\mu_\rho^\nu_\sigma}+\tensor{R}{^\mu_\sigma^\nu_\rho}\right)\, .
\end{split}
\end{align}

It is also possible to apply the RNC to the spin-connection since it depends on the vierbeins and the Christoffel connection. It boils down to using a FS gauge for the spin-connection as if it were a regular gauge field,
\begin{align}
\begin{split}
\omega_\mu=&-\frac{1}{8}\gamma^{\alpha_1}(x_0)\gamma^{\alpha_2}(x_0) R_{\mu\nu\alpha_1\alpha_2}(x_0)\,y^\nu-\frac{1}{12}\gamma^{\alpha_1}(x_0)\gamma^{\alpha_2}(x_0)(\n_\alpha R_{\mu\nu\alpha_1\alpha_2})(x_0)\,y^\nu y^\alpha\\
&-\frac{1}{32}\gamma^{\alpha_1}(x_0)\gamma^{\alpha_2}(x_0)\left( (\n_{\alpha\beta} R_{\mu\nu\alpha_1\alpha_2})+\frac{1}{3}\tensor{R}{_\alpha_\mu^\lambda_\beta}R_{\lambda\nu\alpha_1\alpha_2} \right)(x_0) \,y^\alpha y^\beta y^\nu+\mathcal{O}(y^4)\, .
\label{omegaRNC}
\end{split}
\end{align}
Note that the partial derivatives of $\omega$ only apply on $y$, since the Dirac matrices are at $x_0$ they commute with $\pa=\frac{\pa}{\pa y}$.

\paragraph{Example: $\dc_\mu\dc_\nu\dc_\rho\dc_\sigma$}

Expanding open covariant derivatives in terms of the Christoffel connection, spin-connection and gauge connection and their derivatives can become a heavy computation very quickly. Then forming the covariant quantities adds to the complexity of the task. Using the RNC and the FS gauge drastically simplify this task.

Using Eq.~\eqref{VFSgauge}, we can express the partial derivatives of $V$ in terms of field strengths, and likewise for the spin-connection using Eq.~\eqref{omegaRNC}.

If we denote $X=V+\omega$ for simplicity, we have,
\begin{align}
\begin{split}
\dc_\mu\dc_\nu\dc_\rho\dc_\sigma&=\dc_\mu\dc_\nu\dc_\rho X_\sigma\\
&=\dc_\mu\dc_\nu\left((\pa_\rho X_\sigma)-\Gamma^\lambda_{\rho\sigma}X_\lambda+X_\rho X_\sigma\right)\\
&=(\pa_{\mu\nu\rho}X_\sigma)+(\pa_\mu X_\nu)(\pa_\rho X_\sigma)-(\pa_\mu\Gamma^\lambda_{\nu\rho})(\pa_\lambda X_\sigma)-(\pa_\mu\Gamma^\lambda_{\nu\sigma})(\pa_\rho X_\lambda)\\
&+\left.\bigg[ (\pa_\mu X_\rho)(\pa_\nu X_\sigma)-(\pa_\mu\Gamma^\lambda_{\rho\sigma})(\pa_\nu X_\lambda) \bigg]\right|_{\text{sym }\mu\leftrightarrow\nu}\, .
\end{split}
\end{align}
We can then explicit $X=V+\omega$ and use Eqs.~\eqref{VFSgauge}, \eqref{ChristoffelRNC} and \eqref{omegaRNC}  to form the covariant quantities.

Finally we obtain the 4 open covariant derivatives $\dc_\mu\dc_\nu\dc_\rho\dc_\sigma$ in a covariant form. Obviously, $\dc_\mu\dc_\nu\dc_\rho\dc_\sigma$ is not covariant by itself, but all the contributions to this operators that arise in the expansion combine together so that the result is covariant. The use of RNC and FS gauge is merely a shortcut to get to the final covariant form.

\bibliographystyle{JHEP}
\bibliography{biblio}
\end{document}